\documentclass[sn-mathphys]{sn-jnl}
\usepackage{bm,lscape,makecell,amsmath,amsfonts,amssymb,subcaption,hyperref}

\jyear{2021}%

\begin{document}

\title[ ]{COVID-19 Regional Waves and Spread Risk Assessment through the Analysis of the Initial Outbreak in Guatemala}

\author*[1]{\fnm{Juan A.} \sur{Ponciano}}\email{japonciano@ecfm.usac.edu.gt}

\author[1]{\fnm{Juan D.} \sur{Chang}}

\author[2]{\fnm{Mariela} \sur{Abdalah}}

\author[2]{\fnm{Kevin} \sur{Facey}}

\author[3]{\fnm{Jos\'e M.} \sur{Ponciano}}

\affil*[1]{\orgdiv{Instituto de Investigaci\'on en Ciencias Físicas y Matemáticas}, \orgname{Universidad San Carlos de Guatemala}, \orgaddress{\state{Guatemala City}, \country{Guatemala}}}

\affil[2]{\orgdiv{National Advanced Computing Laboratory}, \orgname{National High Technology Center}, \orgaddress{\state{San Jos\'e}, \country{Costa Rica}}}

\affil[3]{\orgdiv{Department of Biology}, \orgname{University of Florida}, \orgaddress{\city{Gainesville}, \postcode{FL 32611}, \state{Florida}, \country{USA}}}


\abstract{The initial surge of the COVID-19 pandemic hit Guatemala on March 2020. On a country scale, the epidemic has undergone a fairly well-known and distinguishable initial phase, reaching its peak on mid July 2020. However, the detailed picture is more involved and reflects inter-regional variations in the epidemic dynamics, presumably grounded on socio-demographic, connectivity, and human mobility  factors. Classifying the regional epidemic curves and identifying the major hubs of regional COVID-19 spread can contribute towards defining an evidence-based risk map for future outbreaks of infectious diseases with similar transmissibility properties. In this work, we make a regional wave decomposition of the initial epidemic phase registered in Guatemala, and we use the Richards phenomenological model alongside multivariate ordination techniques of its estimated model parameters to draw a countrywide picture of the first epidemiological wave. By exploring similarities in the model space parameters, we traced routes for the disease spread across the country. We evaluated how well the proposed classification can help to define a regional risk hierarchy comprising early stage focal points, major hubs, and secondary regions of epidemic progression.}


\maketitle

\section{Introduction}\label{sec:intro}

The novel coronavirus (SARS-CoV-2) outbreak has left a worldwide spatial and temporal progression landscape of a pervasive communicable disease \cite{whoDashboard}. The scattered historical epidemic data represents a tremendous volume of information about an explosive viral infection among individuals who, at an earlier stage, were mostly uninformed and unaware about the potential harm of the virus. Our focus is on the epidemic development in countries with little to no resources to implement a strong epidemiological response. Although in these countries the epidemic information has been unevenly collected and presents several limitations, here we recognize its potential value to explore the course of this epidemic and of future ones.  

Shortly after the pandemic started, a large amount of scientific researchers from many disciplines began to study the collected data and raise the discussion about issues concerning the viral epidemiology and the evolution of the pandemic to a thorough level of analysis  \cite{yang2020modified, yan2020interpretable, chatterjee2020147, Bertozzi16732, shakeel2021covid}.
Despite the reach of detailed studies that make use of modern human mobility information/databases, assessing the spread risk of COVID-19 in countries where such information is not available remains a challenging goal. Even with data from GPS technology from mobile phones, assessing spread risks has proven to be puzzling given the high complexity of the structure of modern social systems. Practical and theoretical challenges to characterize epidemic growth and the fast adaptive evolution of the virus in the host population are all factors that make of accurate risk assessments a difficult target. If these factors are carefully studied and comprehended, a better understanding of the infection behavior could be possible. For instance, \cite{chang2021mobility,acuna2020108370,lopez2020,jia2020population} show that an in-depth risk-assessment of spatial spread modeling could be made feasible by solving the problem of capturing human behavioral change. Furthermore, characterizing human mobility in the initial stage of the epidemic could reveal the most probable propagation routes of the infection in the absence of interventions \cite{Li2021}. Any scientific progress along this line of work would be valuable to practitioners attempting to manage future transmissible airborne epidemics. 

To date, many studies have focused on mid and large-scale population flows to predict geographical distribution of COVID-19 \cite{wu2020nowcasting,jia2020population}. At a domestic scale, the GPS technology in mobile phones offers a fast and accurate tracking of individuals movement. While this tool is extremely valuable, refined nationwide data is not always available nor easily reachable in many countries. The above is particularly true for the Central American region. Hence, in this study, we focused on developing and proposing an evidence-based approach to model the spread risk of an airborne disease outbreak within one developing country. To do that, we characterize the spread progression of COVID-19 only at an early stage. As a focus study we chose the country of origin of three of us: the Latin American country Guatemala. Our modeling strategy can be implemented to any epidemic outbreaks with similar transmissibility to the initial phase of the COVID-19 epidemic, with mild or no interventions.

Solving the puzzle to establish a time line for the spatial spread of COVID-19 initial phase in Guatemala begins with the identification of regional epidemic waves and the proper characterization of their relevant phenomenological features. On the other hand, a spatial risk assessment of this disease relies on identifying key trends present in the COVID-19 regional wave landscape and finding out plausible spatial associations.

In this study we develop a systematic procedure to identify the initial wave from daily incidence curves and to search for suitable analytical approximations from a phenomenological epidemic model.   We show that the Richards Model, a straightforward generalization of the classical Logistic Model \cite{richards1959flexible}, notably approximates a wide range of growth curves configurations, thus provides a convenient tool for wave description.  Our approach to wave classification is rooted in unconstrained ordination \cite{abdi2010principal} of the Richards curves parameter estimates for every geographic locale to look for similarity patterns in the waves properties as embodied in these parameters. We further take advantage of available estimates about the country  inter-regional population flow to combine wave patterns with spatial mobility data for spread risk assessment. 

Our approach on the spatial association of regional outbreaks may prove to be useful to refine epidemic risk strategies, particularly in the context of countries devoid of resources for scientific research and a robust epidemic response infrastructure.  Finally, we also illustrate how the same information that typically is used to provide COVID-19 trackers at National levels without any further insights can be used to generate actionable risk information for this and future epidemics in these countries.

The rest of this paper is organized as follows: In Section \ref{sec:Back} we explain the bases of logistic models and Richards equations. A brief description of related work and the results of our study are exposed and discussed in Section \ref{sec:res}. The final conclusions of our analysis are presented in Section \ref{sec:Final}.
Finally, section \ref{sec:met} describes the followed methodology regarding data treatment, wave identification and classification, parameter estimation, PCA analysis, and spatial risk assessment. 

\section{Background}
\label{sec:Back}
\subsection{Logistic Models and Disease Outbreaks}

There is a vast literature showing that logistic phenomenological models are capable of capturing the main features of growth profiles in single infectious disease outbreaks \cite{ma2014estimating,chowell201666, Pell2018, Brauer2019}. Uncontrolled propagation that leads to exponential curves is the most common assumption at the very beginning of an emerging epidemic. Accordingly, the increase of the number of infected cases $c(t)$ over time is  properly described by the exponential growth model, which states that the per capita rate of change of the total number of cases equals a constant $r$ \cite{chowell201666,viboud2016generalized}. Explicitly, the exponential model is

\begin{equation}
\frac{1}{c(t)}\frac{dc(t)}{dt} = r
\end{equation}

The $r$ parameter sets the timescale of the epidemic progress and is known as the intrinsic growth rate. At the initial stage of an epidemic, $r$ can be used to obtain an estimate of the basic reproduction number $R_0$ \cite{heffernan2005}, the number of secondary infection cases that are derived from one contagious individual.

The central  hypothesis in the exponential model lies on the assumption that, at all times, the per capita contribution to the growth of the epidemic is unaffected by the total number of cases. However, a variety of factors acting as ``density-dependent processes" can affect the evolution of the epidemic, causing the outbreak to slow down and to reach a phase with decaying transmission rate \cite{chowell201666}. The exponential model fails to follow this behavior, which is instead faithfully reproduced by the logistic approach. These models can be conceived as variations of the standard logistic growth model, formulated in terms of two parameters: the generalized growth rate $r$ and the carrying capacity $K$. This second parameter accounts for the total number of cumulative infections over the whole outbreak. The standard logistic model heretofore abbreviated as LM is defined as \cite{ma2014estimating}.

\begin{equation}
\frac{1}{c(t)}\frac{dc(t)}{dt} = r\left(1-\frac{c(t)}{K}\right)
\end{equation}
 
Logistic type models are commonly assumed to be of empirical nature and do not rely on a specific basis of mechanistic assumptions. They are instead rooted on general assumptions about the shape of incidence curves, making use of a few free parameters. However, it can be demonstrated that they approximately reproduce the underlying mechanistic dynamics of the standard SIR model where susceptible individuals are being converted into infected individuals \cite{ma2014estimating}.  In fact, a little known result from theoretical ecology is the work of Dennis (1978) \cite{dennis1978analytical} who showed that anytime a population of interest (here the cumulative number of COVID-19 cases) grows on a supply of a single limiting nutrient in a open system (here the susceptible individuals), the resulting growth is logistic like. Furthermore, in closed systems, the growth of such population in these circumstances is exactly logistic.  Briefly, Dennis' \cite{dennis1978analytical} mathematical argument starts by modeling the process of conversion of nutrient into population growth using a system of two differential equations (one for the nutrient, one for the population of interest) that very much resembles mathematically the standard SIR model.  He then shows that reducing this system of equations to a single equation is possible and importantly, the resulting equation is logistic like and in a special case, exactly logistic.  These results from theoretical ecology bring more relevance to phenomenological models of the logistic type, as they provide equations with less parameters that bear a close relation to epidemic dynamics. Importantly, the work of Dennis \cite{dennis1978analytical} and \cite{ma2014estimating} show that the logistic-like model parameters are themselves the product of parameters carrying information regarding the process of conversion of susceptible into infected. The link of these models with mechanistic processes is also likely to result in stable statistical fits and reliable estimates in data analyses \cite{wu2020generalized}.

\subsection{Richards Model}

The Richards Model \cite{richards1959flexible}, originally introduced for ecological population growth as the ``theta-logistic model'' \cite{sibly2005regulation}, provides a useful generalization of the classical LM. This model includes an extra parameter, resulting in a more flexible growth function and thus, accounting for variations in the transition time from the accelerating to the decelerating phase of the epidemic curve.

In the Richards Model (RM), the differential equation for the cumulative cases $c(t)$ is given by

\begin{equation}\label{RM}
\frac{dc(t)}{dt}=rc(t)\left[ 1- \left( \frac{c(t)}{K} \right)^a \right]
\end{equation}

When $a=1$, RM is reduced to the standard logistic model. On the other hand, when $a\ll1$,  the density-dependent term is lost.

The exact solution to Equation (\ref{RM}) is explicitly given by

\begin{equation}\label{RSol}
c(t)=K\left( 1+a e^{-a r(t-t_c)}\right)^{-1/a}
\end{equation}

where $t_c$ is the time where the second derivative of $c(t)$ vanishes and therefore, corresponds to the inflection point from increasing to decreasing rate of cumulative cases.

Although some of the parameters in the Richards empirical model may not have a straightforward connection with epidemiological concepts, contrary to the widely used compartmental SIR models, Equation (\ref{RSol}) describes the cumulative case data in a surprisingly accurate manner. Besides, some terms in the Richards equation have been found to contain intrinsic relations to the ones of the SIR model (e.g., the exponential term). A drawback of using the Richards model approach is the curve over fitting tendency, for which it is necessary to add some constraints in the modeling process \cite{WANG201212}.  In any case, here we show that using this simple model in the face of incomplete epidemic information is useful to assess the temporal and spatial epidemic risk of COVID-19 and other similar communicable diseases.

\section{Methods}\label{sec:met}

\subsection{Wave Identification and Maximum Likelihood Estimation}

Prior to the fitting process, we estimated the inflection points of the observed epidemic curve, by implementing a computational algorithm to determine the zeros of the variation of daily reports. For each department, we applied a digital filter to the running average data to obtain a smoothed version of the curve behavior. This filter was performed through a Savitzky-Golay filter \cite{Savitzky1964}.   This filter was also employed to identify the location in time of the inflection point from increasing to decreasing rate of cumulative cases, within a wave interval. 

Once the beginning point and the duration of each wave was approximated, we truncated the smoothed version of the cumulative growth curve to a time window, defined by the results of the wave identification in each location. Then, we started the fitting procedure by adjusting the parameters of the Richards curve.

We chose the set of parameters $r, a, K$, and $t_c$ of Equation (\ref{RSol}) that maximized the likelihood function $L(\bm{\theta})$, where $\bm{\theta}$ is the vector of parameters. We used the customary observation error model assumption that, for the $ith$ department, the observed number of cumulative cases $C_i$ followed a normal distribution. The mean of this distribution was given by a deterministic process predicted by the Richards model, that is

\begin{equation}
C_i\sim \mathcal{N}(\mu=K_i\left( 1+a_i e^{-a_i r(t-t_{c_i})}\right)^{-1/a_i},\sigma^2), \qquad i=1,2,\cdots, 22
\end{equation}

Our likelihood optimization routine followed an embedded algorithm process.  Initially, we optimized the $\log$-likelihood $\log(L(\bm{\theta}))$ via the Nelder-Mead numerical optimization method but this strategy routinely got trapped in local optima.  To solve this problem, we embedded a Nelder-Mead search into each Monte Carlo step of a Simulated Annealing Algorithm (SAA). This strategy not only proved to do better explorations of the parameter space in the $(K, t_c)$ dimensions but also systematically improved on Nelder-Mead searches that got trapped in local minima. To get reliable search results, we implemented random changes in $K$ and $t_c$ values in the neighboring states set of the Markov chain.  We performed 300 Monte Carlo steps to generate sample states, and an annealing schedule of 50 different temperatures.  This strategy was sufficient to warrant convergence at a prescribed tolerance (see R code in GitHub:  
\url{https://github.com/jmponciano/Ponciano-JA-etal-COVID-Guatemala}).

\subsection{Dimensional Reduction and Wave Classification}

The wave classification analysis was carried out with Principal Component Analysis ($PCA$) aimed at reducing the $4$-dimensional parameter space into a lower dimensional space that revealed patterns of similarity in the fitted curves. This reduction was achieved by defining an orthogonal projection of the wave parameters space onto a lower dimensional linear space, such that the variance of the projected points is maximized \cite{johnson2014applied}. The reduced space obtained via $PCA$ was generated by the eigenvectors of the parameters covariance matrix corresponding to the two largest eigenvalues. We call this reduced space the $PCA$ parameters subspace.  The original data matrix consisted of a $22 \times 4$  data matrix $X$ corresponding respectively to 22 regional waves, each one described by 4 parameters.  Because $PCA$ is known to be unreliable when the scale of the ordination variables varies widely, we first standardized all parameter estimates to the same scale in the matrix $X$. We then computed the empirical covariance matrix $S=\rm{Cov}(X)$ of the wave parameters. Next, we considered a 2D subspace, which is determined by the $S$ eigenvectors with the two largest eigenvalues. By construction, the orthogonal projection of the data points into this subspace maximizes the variance of the projected points. We used our own algorithm for the eigen-decomposition of $S$, as well as the one for computing the scores $z_i^{(1)}$ and $z_i^{(2)}$, corresponding to the projections of the data into the first principal components $PC1$ and $PC2$, respectively. By considering individually the one-dimensional spaces defined by $PC1$ and $PC2$, and computing the euclidean separations among the corresponding principal components scores, i.e., $d_{ij}=\vert z_{i}^{\alpha}-z_{j}^{\alpha}\vert$ for $\alpha=1,2$, we established different classifications of the regional waves. This was achieved by resorting to the hierarchical clustering technique within the Cluster package in R.
 
\subsection{Mobility Matrix and Spatial Association}

Classifying the waves by region allowed us to establish a timeline for the spatial spread of COVID-19 virus in Guatemala.  This timeline allowed in turn exploring the spatial association among different regions by constructing a mobility matrix from internal migratory flows. The migratory statistics at departmental level were retrieved from the National Institute of Statistics \cite{INE2018}. In this study, we considered the data related to the amount of workers and students in every municipality that traveled to another department daily, as reported by the last population census carried out in the country in 2018 \cite{INE2018}. As the information was reported with a municipality granularity level, we aggregated the data to determine the quantities per department.

Following Lai et al.\cite{Lai2020}, we define for each department a risk of virus importation, $\rho$, as the number of travelers received divided by the total number of persons leaving the departments that have a high risk of experiencing international infection during one day. We used this information to complement the spatial analysis association of growth curves.

All the algorithms that are mentioned in this section were programmed in R.

\section{Results and Discussion}

\label{sec:res}
 
The Richards model has been widely employed for the general description of several multiple-wave epidemic events worldwide \cite{Shanafelt2018, sanna2018spatial, Pell2018, wu2020generalized, Zuhairoh2020,WANG201212, Macedo2021}. Also, it has been useful to estimate quantities of epidemiological relevance, such as the initial growth rate and the inflection point of an outbreak in a given location \cite{ma2014estimating,hsieh2010epidemic}.

Some authors have applied the Principal Component Analysis technique to select the main COVID-19 severity predictors \cite{Markovic2021} or to evaluate the impact of containment strategies by public polices classification \cite{Chen2021}. In terms of model parameter estimation, Agarwal et al. \cite{Agarwal2021} utilized a PCA to obtain the fundamental parameters of a complex BHRP network model of the second wave of COVID-19. Mahmoudi et al. \cite{Mahmoudi2021} performed a PCA to classify high-risk countries according to their respective numbers of patients and deaths. Another study presents a similar analysis of French cities, but employs the PCA technique to categorize departments in accordance with multidimensional variables such as epidemic indicators and predominant age. Among other findings, the authors reported a relationship between department mortality and the time mobility restrictions were adopted in that specific location \cite{Gaudart2021}. To the best of our knowledge, no studies about COVID-19 wave classification have been published.   
 
Here, we employed the official surveillance data of daily reported COVID-19 cases in Guatemala to carry a wave classification analysis based on both, a fit of the Richards model and a subsequent principal component analyses of the parameter estimates for each region. The only information needed was retrieved from the COVID-19 web platform, managed by the national Ministry of Public Health (MSPAS) \cite{MSPAS} and consisted of the time series of the daily registered cases, starting from March 13th, 2020 onwards. The infected cases were registered at the onset time of the disease symptoms. In this research, we limited the wave identification analysis to a departmental level (Guatemala is organized into 22 different Departments equivalent to states in the USA, and within each department there are multiple ``municipalities'' akin to counties in the USA). In total, there are 340 municipalities in the country. As the data granularity is based on municipalities (the second-level administrative divisions of the country as mentioned above) we pooled the data from all municipalities per department in the country to obtain the cumulative values for the 22 departments in Guatemala.

For the wave identification process, we first calculated the moving average of the daily cases with a 14 day window. For model fitting using the Richards equation, we made use of the time series of cumulative cases of each department (see Methods). The time windows used for the analyses started 65 days after the first reported case in the country, in order to reduce the noisy regime with strong fluctuations of the very early outbreak. For each department, the wave identification was performed by fitting the first phase of the cumulative infected to the Richards Model via maximum likelihood (see Methods). 

\begin{figure}
     \centering
     \begin{subfigure}[h]{0.48\textwidth}
         \centering
         \includegraphics[width=\textwidth]{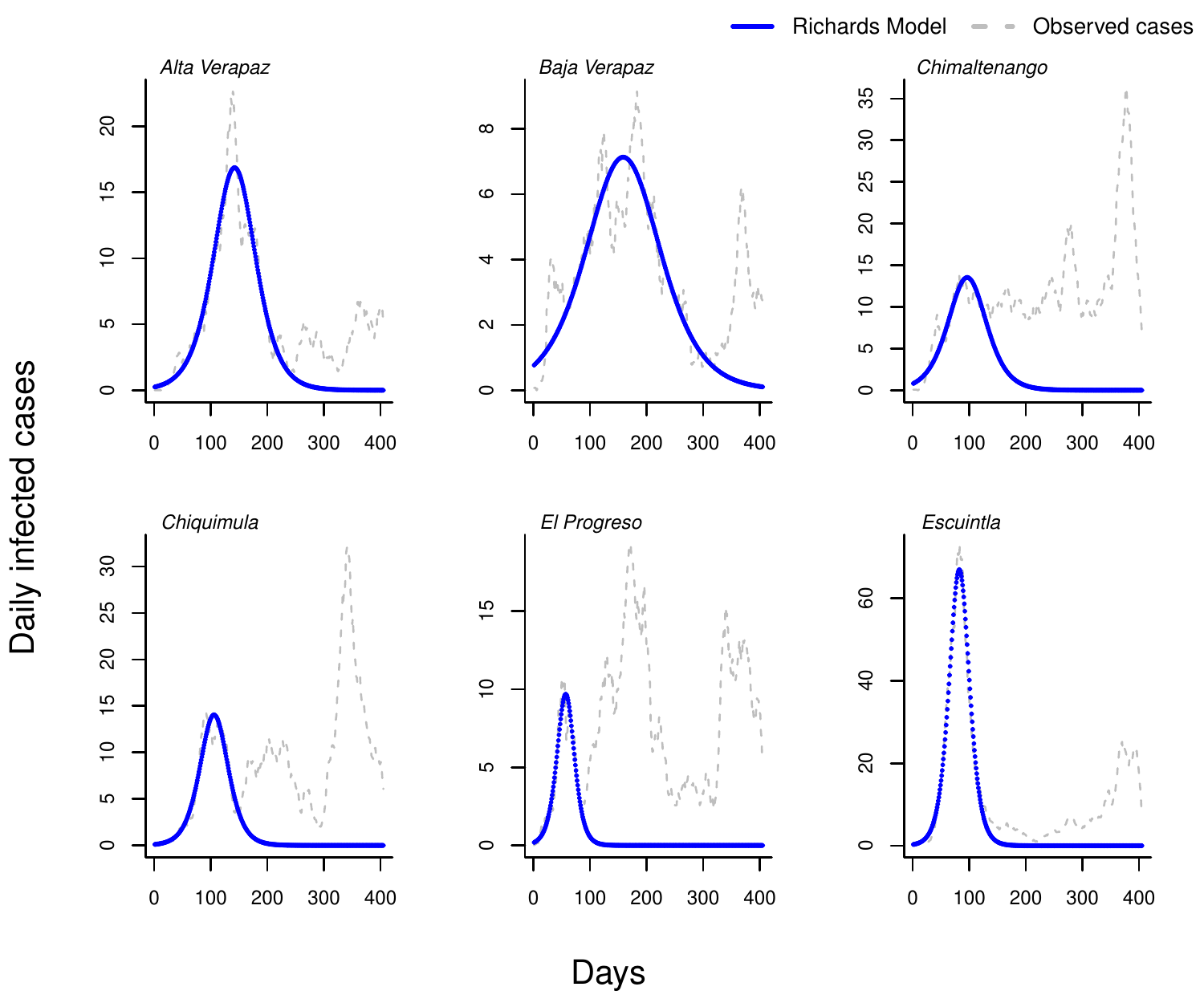}
         \label{fig:1a}
     \end{subfigure}
     ~ 
     \begin{subfigure}[h]{0.48\textwidth}
         \centering
         \includegraphics[width=\textwidth]{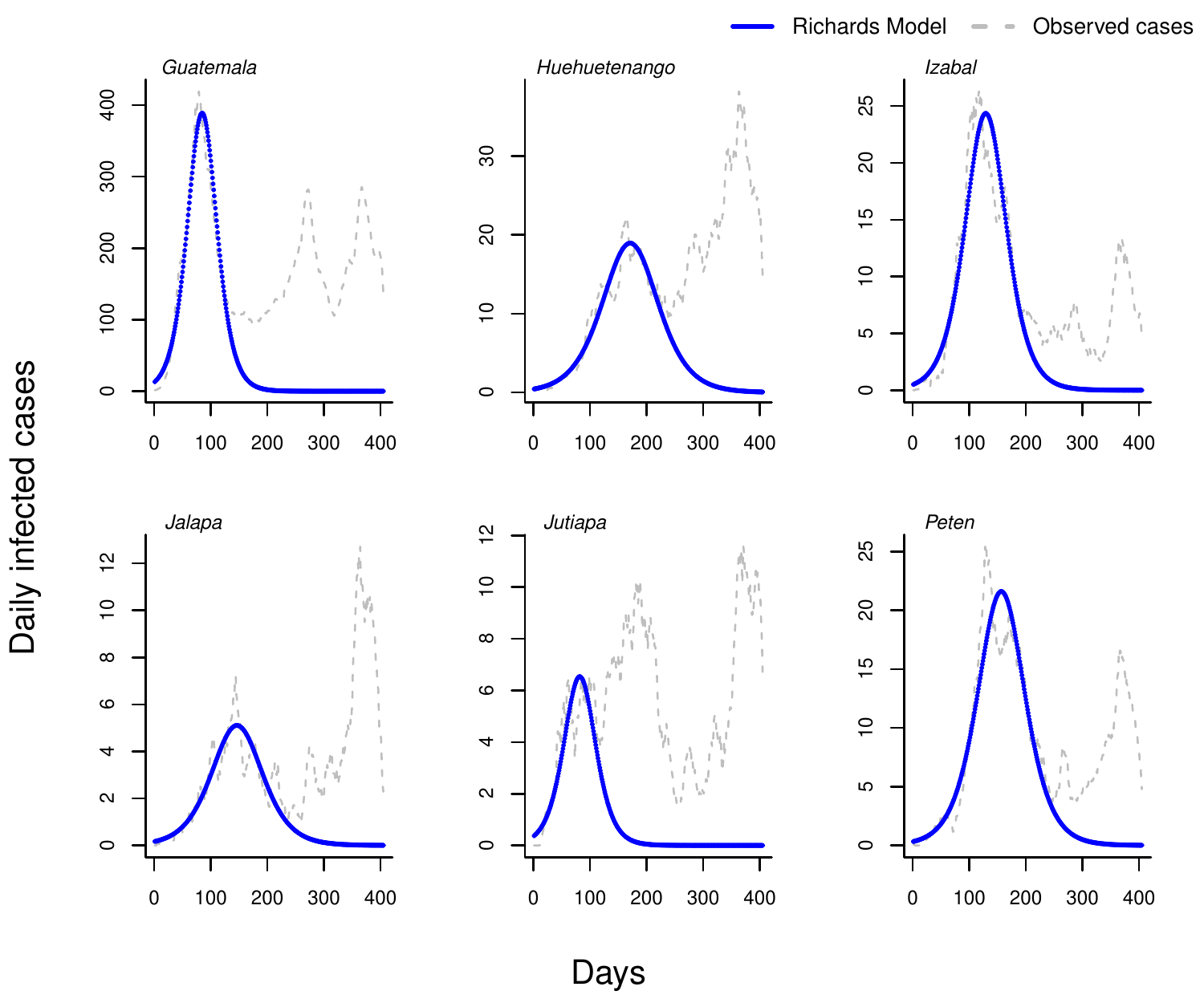}
         \label{fig:1b}
     \end{subfigure}
     \begin{subfigure}[h]{0.48\textwidth}
         \centering
         \includegraphics[width=\textwidth]{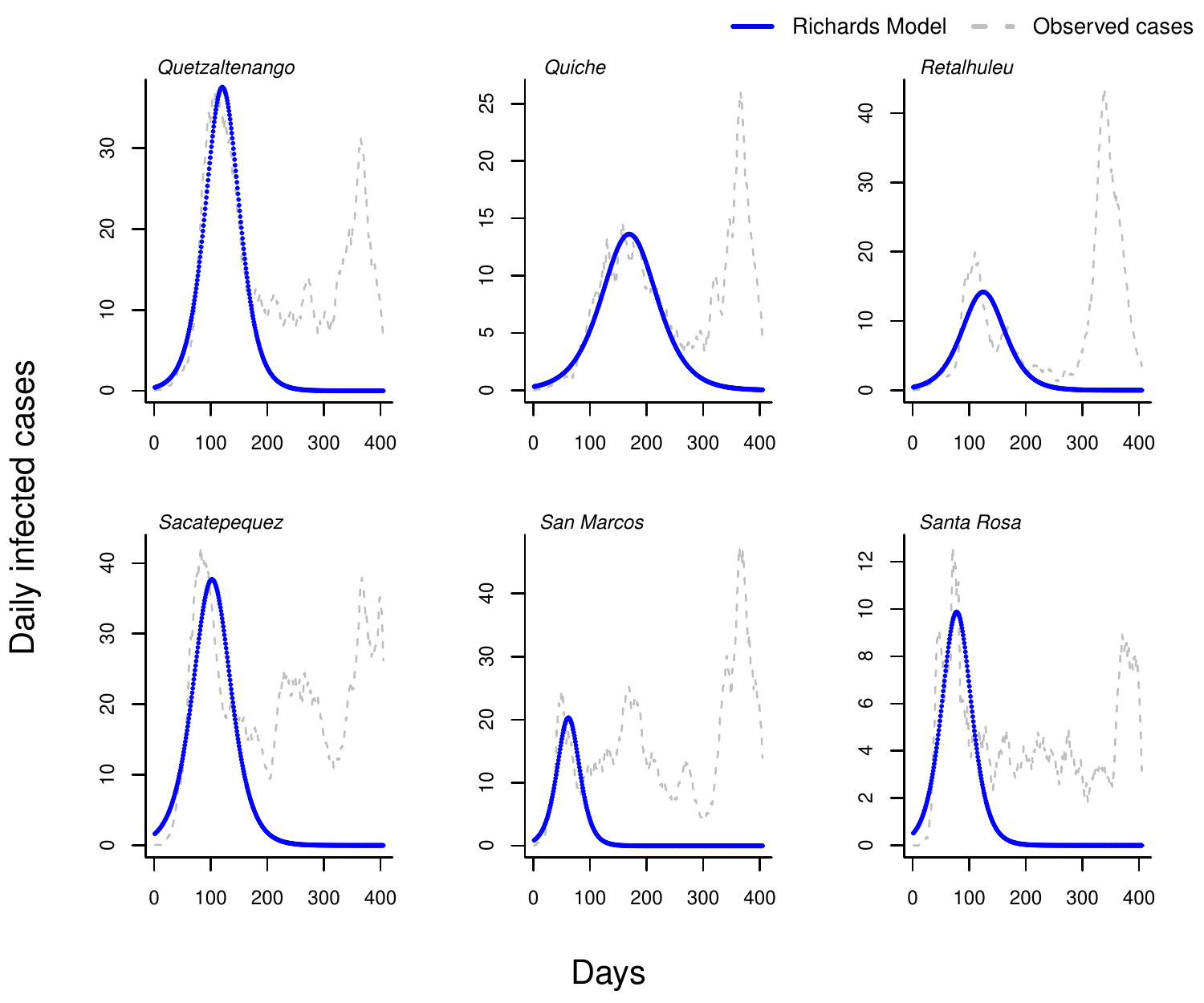}
         \label{fig:1c}
     \end{subfigure}
     ~ 
    \begin{subfigure}[h]{0.48\textwidth}
         \centering
         \includegraphics[width=\textwidth]{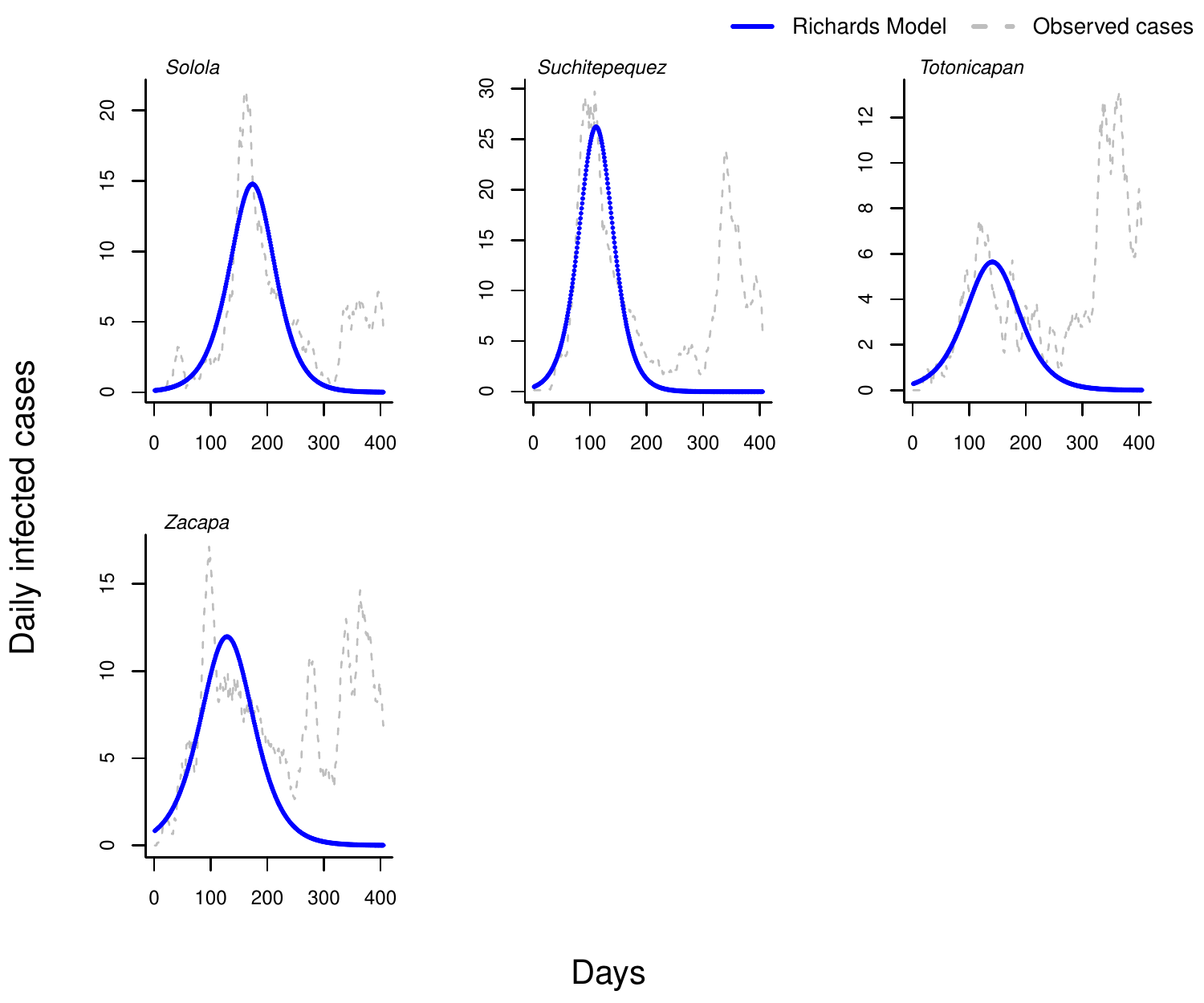}
         \label{fig:1d}
    \end{subfigure}
    
\caption{Richards model fit for the observed COVID-19 daily infected cases in the 22 departments of Guatemala}\label{Fig:1}
\end{figure}

The results of fitting the Richards Model to the first wave of registered COVID-19 cases are displayed in Figure \ref{Fig:1}. Table \ref{table:1} contains a summary  for the maximum likelihood model parameters estimates for each department. In general, the inflection points of the regional first waves related to the Richards model are found from 57 to 174 days, starting from day 65 after the first reported case in the country.

\begin{table}[ht]
 \centering
 \caption{Maximum likelihood parameter estimates of the Richards model for COVID-19 cases in the 22 departments of Guatemala}\label{table:1}
 \begin{tabular}{|l|r|c|c|c|c|}
 \hline
 Department & Population \cite{INE2018} &$a$ & Growth rate, $r$ & \makecell{Carrying \\ capacity, $K$} & \makecell{Inflection \\ point, $t_c$}\\
 \hline
 Alta Verapaz & 1215038 &1.028& 0.040 & 1685 & 142\\
 Baja Verapaz & 299476 &1.016 & 0.023 & 1258 & 159\\
 Chimaltenango & 615776 & 1.010 & 0.044 & 1233 & 96\\
 Chiquimula & 415063 &1.008 & 0.060 & 941& 106\\
 El Progreso & 176632 &1.062 & 0.099 & 400 & 57\\
 Escuintla  & 733181 & 1.005 & 0.084 & 3195 & 82 \\
 Guatemala & 3015081 &1.011 & 0.057 & 27230 & 85 \\
 Huehuetenango & 1170669 &1.006 & 0.030 & 2509 & 171\\
 Izabal &408688  & 1.006 & 0.041 & 2374 & 129\\
 Jalapa & 342923 &1.009 & 0.033 & 618 & 146 \\
 Jutiapa & 488395 &1.029 & 0.053 & 495 & 82  \\
 Pet\'en & 545600 &1.004 & 0.036 & 2415 & 156\\
 Quezaltenango & 799101 & 1.004 & 0.049 & 3048 & 120\\
 Quich\'e & 949262 &1.004 & 0.030 & 1783 & 169\\
 Retalhuleu  & 326828 &1.003 & 0.039 & 1448 & 125\\
 Sacatepequez & 330469 &1.010 & 0.045 & 3374 & 102\\
 San Marcos & 1032277 &1.022 & 0.075 & 1087 & 62\\
 Santa Rosa & 396607  & 1.020 & 0.057 & 697 & 78 \\
 Solol\'a & 421583 &1.089 & 0.038 & 1598 & 174\\
 Suchitepequez & 554695 &1.004 & 0.049 &2142& 111 \\
 Totonicap\'an & 418569 &1.014 & 0.031 & 723 & 141\\
 Zacapa &245374 &1.013 & 0.032 & 1518 & 129\\
 \hline
 \end{tabular}
 \end{table}

The resulting waves can be represented by four parameter estimates from the Richards model.  The points in a $4-$dimensional space defined by these estimates were taken as the basis for the PCA and calculation of the wave parameter space onto the principal subspace.

\begin{figure}
 \centering
    \begin{subfigure}[h]{0.75\textwidth}
    \centering
    \includegraphics[width=0.9\textwidth]{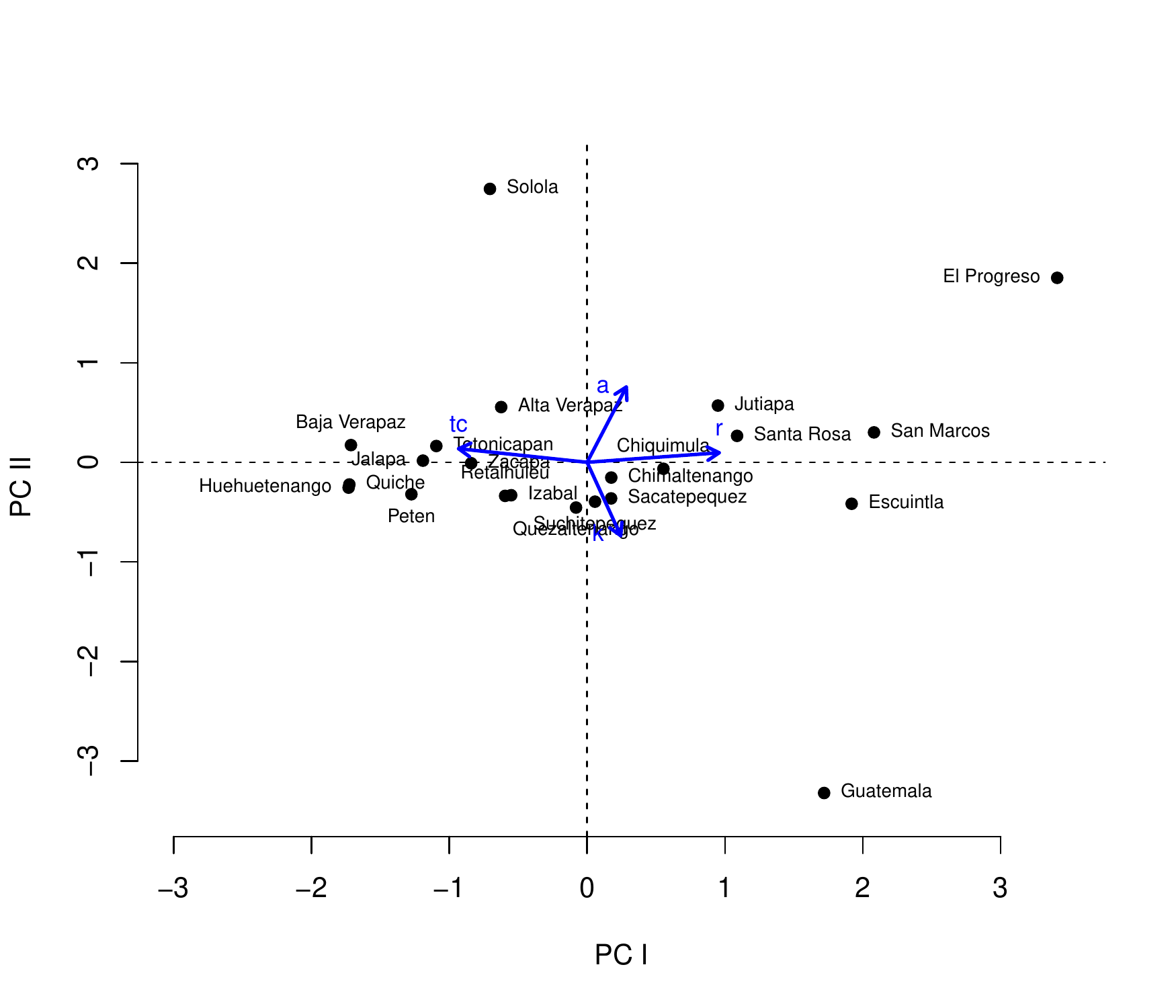}
    \subcaption{}
    \label{fig:2a}
 \end{subfigure}
 ~
 \begin{subfigure}[h]{0.75\textwidth}
 \centering
 \includegraphics[width=0.9\textwidth]{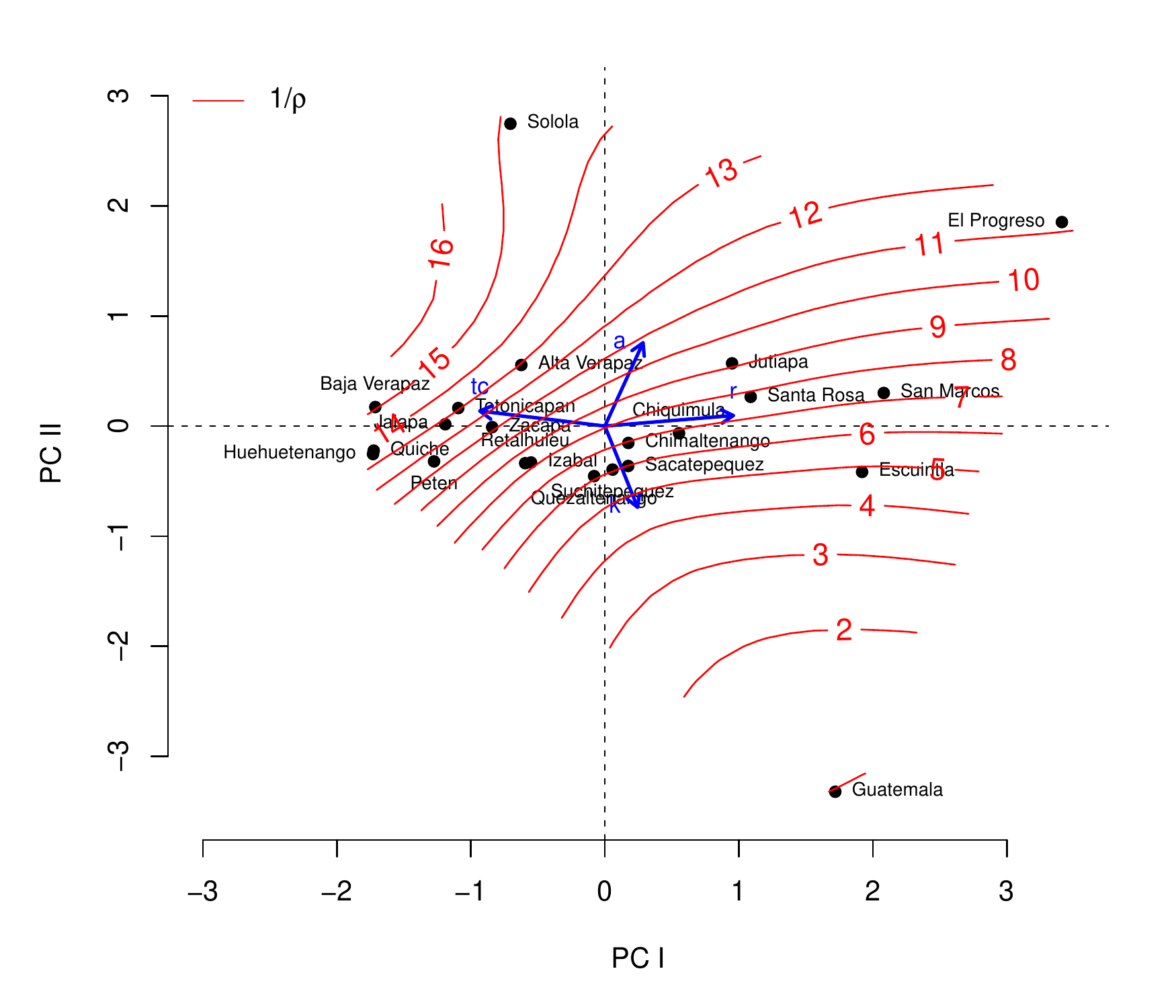}
 \subcaption{}
 \label{fig:2b}
 \end{subfigure}
 \caption{Obtained principal component space (a) and relation of the obtained principal component space with the inverse virus importation risk (b)}
\end{figure}

We found that the first principal component, $PC1$, explains $48\%$ of the overall variability, while $PC2$ accounts for $29\%$ of it. This means that  $77\%$ of the variability is related just to $PC1$ and $PC2$; hence, we reduced the variability analysis to the space defined by these two components. The projection of the wave parameter data onto the $2D$ principal subspace is displayed in Figure \ref{fig:2a} along with the projections of the vector basis given by $r,a,t_c$ and $K$. Examining the correlation matrix between the model parameters and the principal components displayed in table \ref{table:2} as well as Figure \ref{fig:2a} it can be seen that the first principal component scores mainly capture similarities in the peak position ($t_c$) and hence the time span of the regional waves. On the other hand, from Figure \ref{fig:2a} we can see that the second principal component dimension gives more information about the carrying capacity $k$ and the $a$ exponent parameter.

\label{fig:2}

\begin{figure*}
 \includegraphics[width=\textwidth]{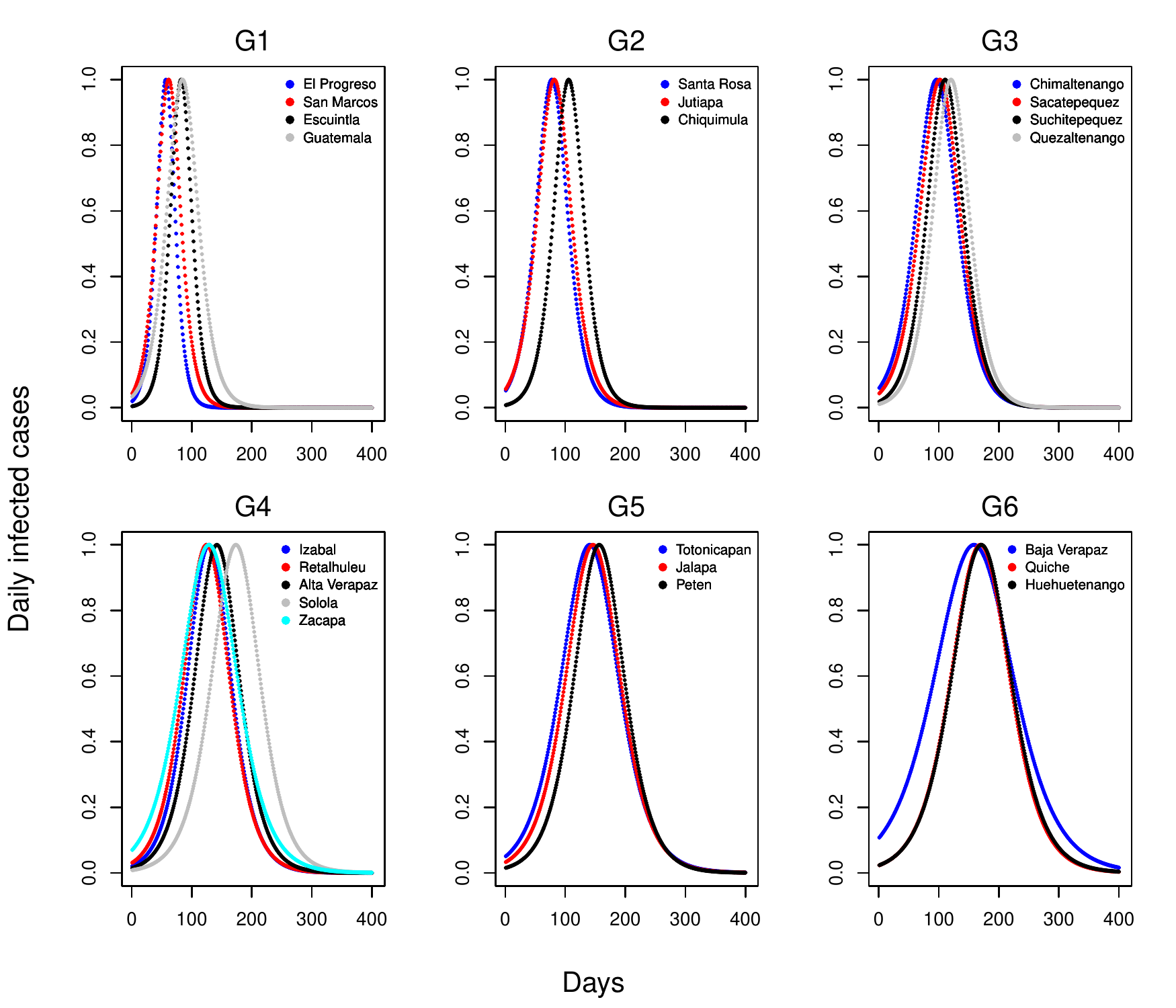}
\caption{Classification of COVID-19 waves by the first principal component scores in the 22 departments of Guatemala }\label{fig:3}
\end{figure*}

Figure \ref{fig:3} illustrates the results of the classification of regional waves by using euclidean distances, according to the $PCI$ score associated to each wave. We found six different groups of waves that were labeled from $G1$ to $G6$. In this figure, daily infected cases for each department have been normalized by their maximum value. 

\begin{table}[ht]
 \centering
 \caption{Correlations of Richards model parameters with principal components}\label{table:2}
 \begin{tabular}{|l|c|c|c|c|}
 \hline
 &PC1&PC2&PC3&PC4\\
 \hline
 $a$&0.28 &0.76 &- 0.59& 0.06\\
 $r$&0.96 &0.06 &0.11 & -0.25\\
 $K$&0.24 &-0.74 &-0.63 & -0.01\\
 $t_c$&-0.93 &0.14 &-0.23 &-0.24 \\
\hline
 \end{tabular}
 \end{table}

Wave classification results using Principal Component scores allowed us to propose associations in the spatial distribution of the infection through the following scenario:

First, the group of regional waves with the lowest mean inflection point $t_c$ seems closely related to hot spots of virus importation (groups $G1-G2$). Second, departments characterized by mid $t_c$ values have presumably high connectivity to those hot spots (groups $G3-G4$). Also, departments with significant population mobility (group $G3$) can act as secondary sources for regional epidemic spread, boosting in that way the epidemic course of rural and more isolated regions that belong to groups $G5$ and $G6$. Finally, once the outbreak gets triggered, the interplay of mobility patterns and population density shapes the main traits of the regional growth curves.

The proposed scenario above strongly suggests that geographical connectivity is encoded in the global picture of the regional COVID-19 wave landscape. Following this rationale, the local growth curves characterization can help to single out the main elements that are necessary to depict a spread risk map of the disease. Our investigation of the spatial associations in the context of the COVID-19 epidemic and the estimation of a risk map from these also suggest that this idea can be extended to model the spread risk of any airborne infectious disease with similar epidemiology.

\begin{figure}
     \centering
     \begin{subfigure}[h]{\textwidth}
         \centering
         \includegraphics[width=0.48\textwidth]{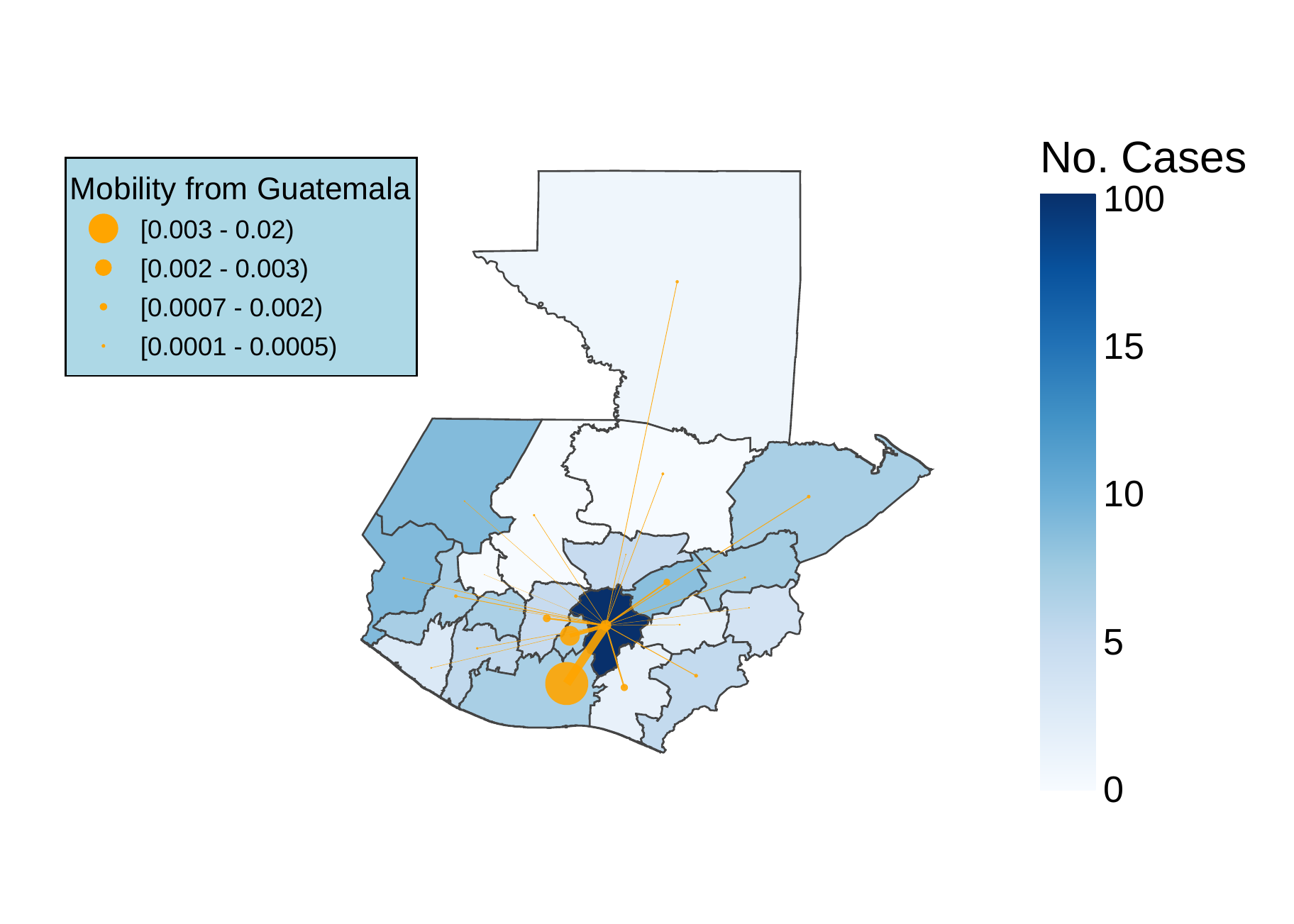}
         \hfill
         \includegraphics[width=0.48\textwidth]{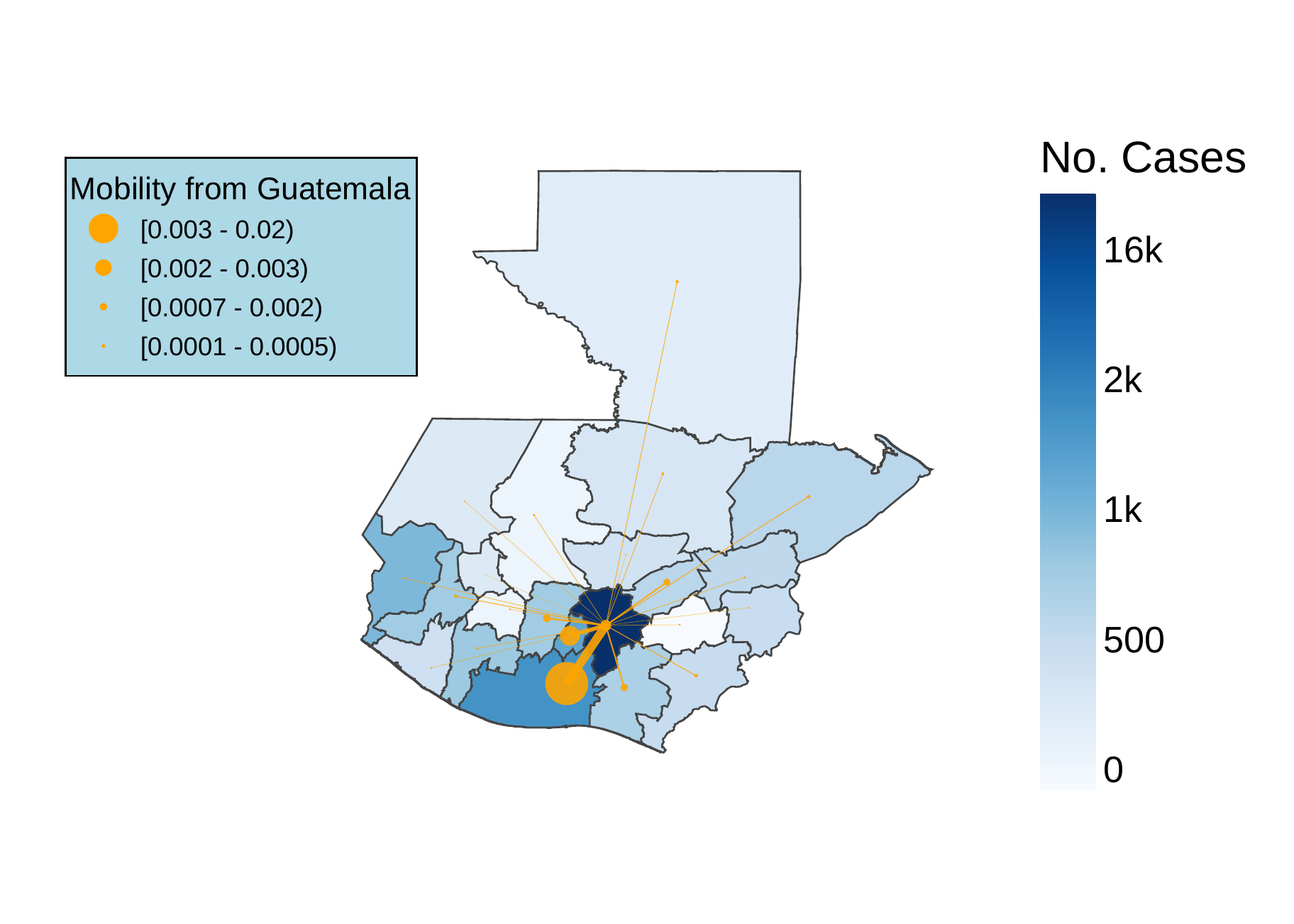}
         \subcaption{}
         \label{fig:4a}
     \end{subfigure}

     \begin{subfigure}[h]{\textwidth}
         \centering
         \includegraphics[width=0.48\textwidth]{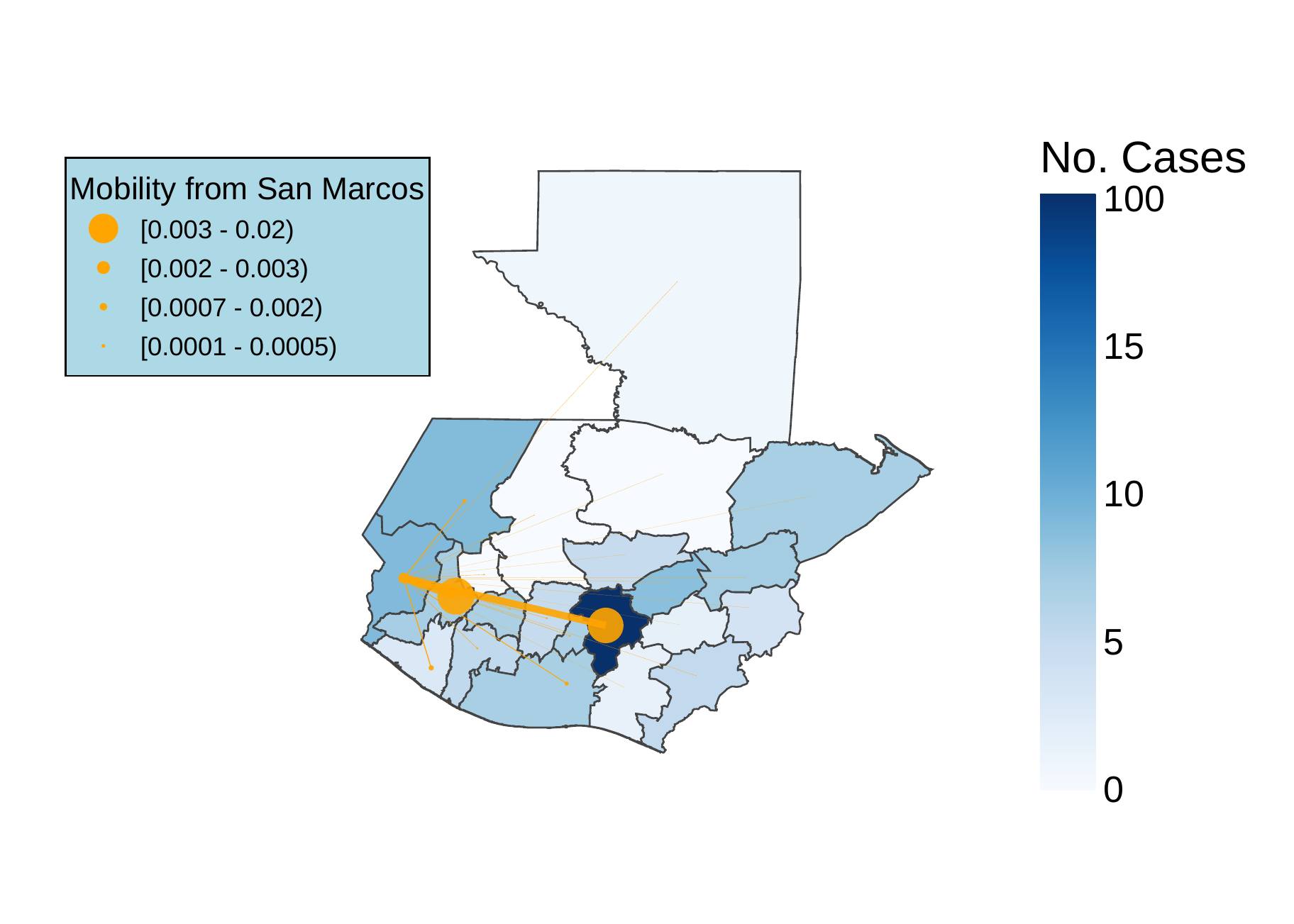}
         \hfill
         \includegraphics[width=0.48\textwidth]{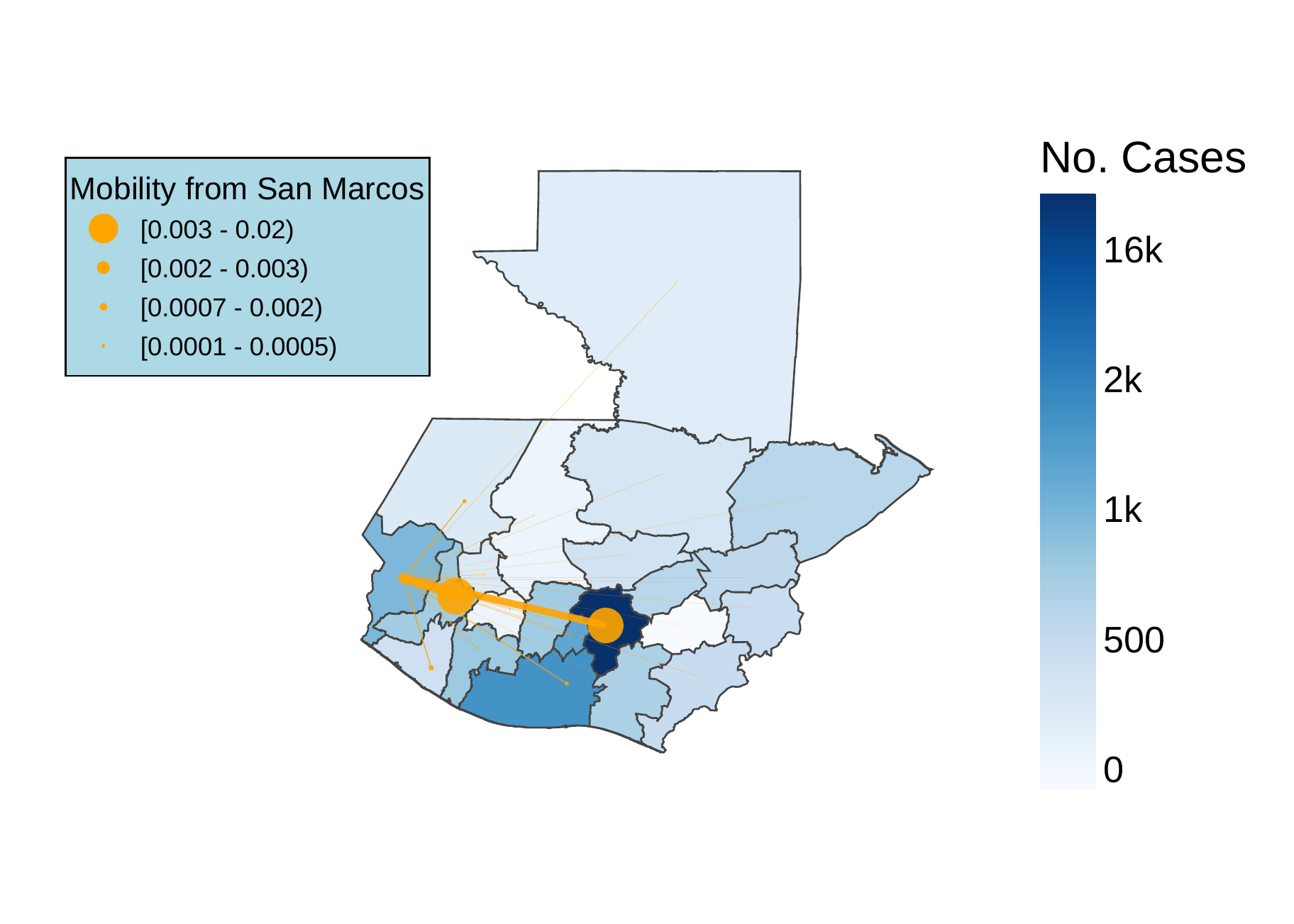}
         \subcaption{}
         \label{fig:4b}
     \end{subfigure}

    \begin{subfigure}[h]{\textwidth}
         \centering
         \includegraphics[width=0.48\textwidth]{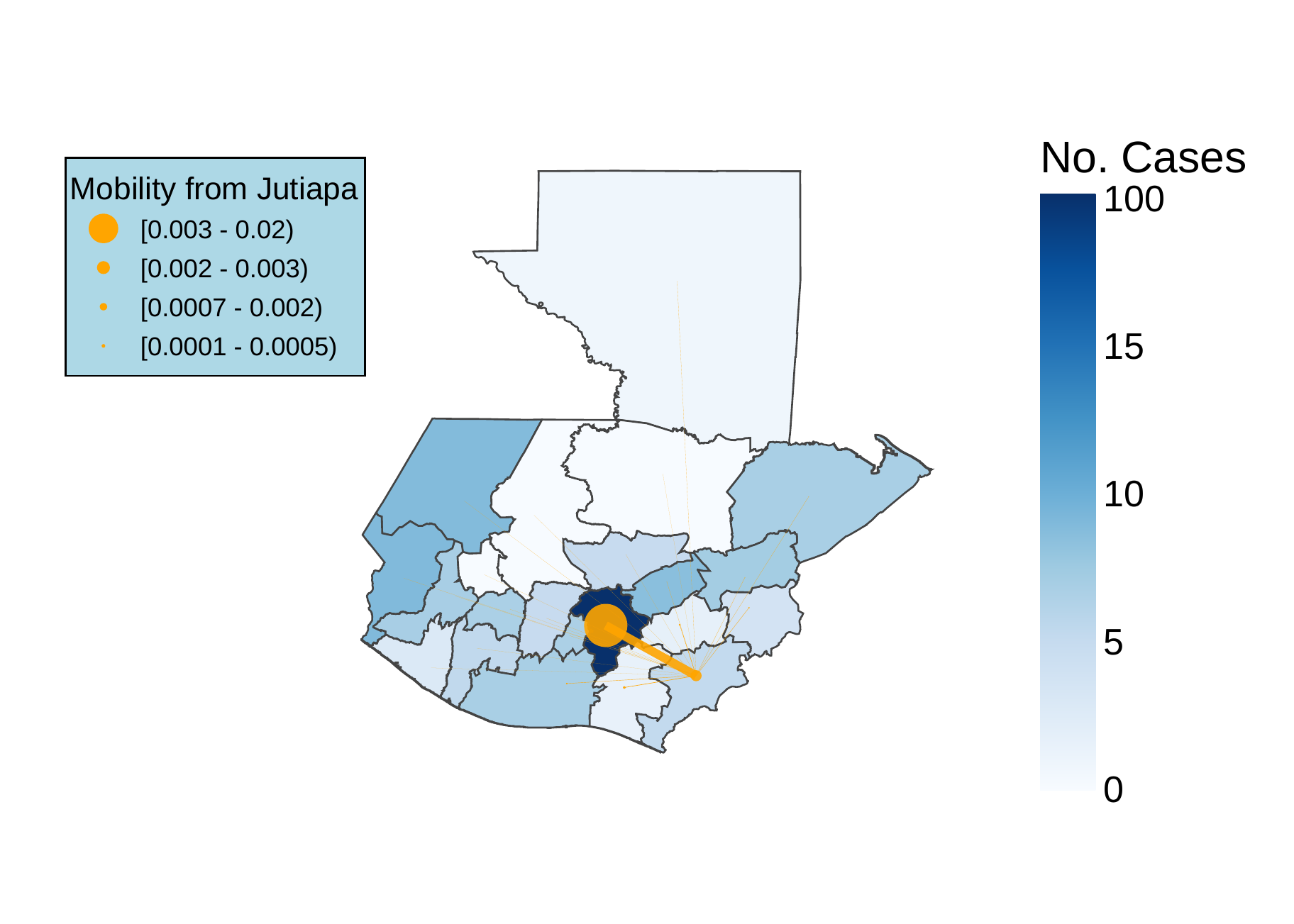}
         \hfill
         \includegraphics[width=0.48\textwidth]{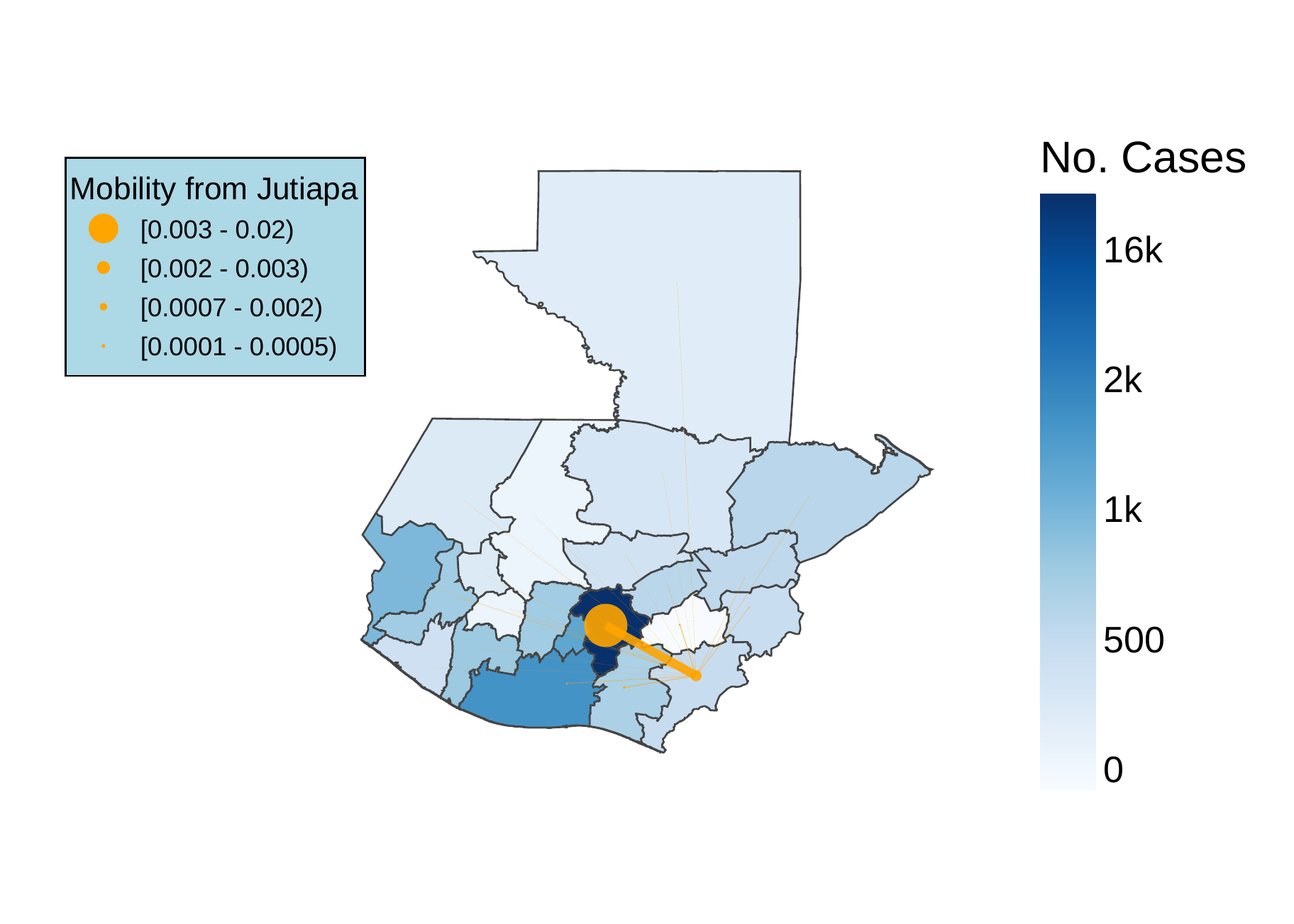}
         \subcaption{}
         \label{fig:4c}
     \end{subfigure}

\caption{Cumulative cases and mobility map for the departments of (a) Guatemala, (b) San Marcos, and (c) Jutiapa. For each locality, the left and right maps correspond, respectively, to the beginning and the peak of the initial wave. }\label{fig:4}
\end{figure}

We compared our findings with available information on areas that are most exposed to virus importation, along with regional mobility indicators. In Guatemala, the main international land migratory flows occurs across the southeastern border with Mexico and the southwestern border with El Salvador. Those areas belong to the departments of San Marcos and Jutiapa, and are considered high-risk regions. In addition, the highly populated city of Guatemala, located in the Guatemala department, receives the highest volume of international commercial flights. It is thus the most vulnerable region to virus importation from an international source.  These three high-risk departments are sorted in groups $G1$ and $G2$, according to the $PCI$ scores in Figure \ref{fig:3}. Also, Figure \ref{fig:4} shows a mobility map that represents risk in three disease epicentres; namely, the departments of Guatemala, San Marcos, and Jutiapa. The figure also displays the information about the cumulative cases per department at two different moments of the initial outbreak: the beginning and the peak of the first wave.

\begin{figure}
    \centering
    \begin{subfigure}[h]{0.8\textwidth}
         \centering
         \includegraphics[width=\textwidth]{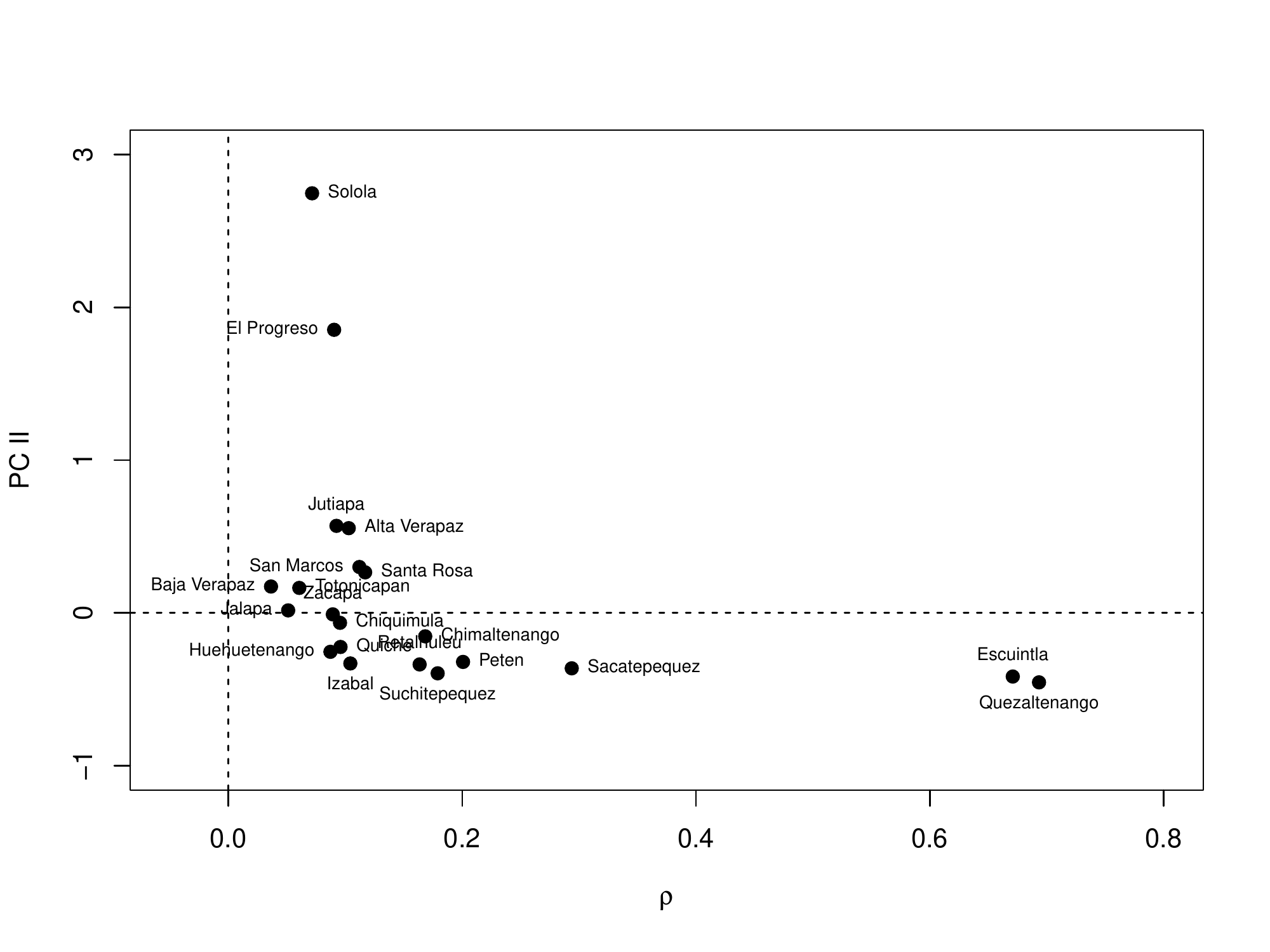}
         \subcaption{}
         \label{fig:5a}
     \end{subfigure}
     ~
     \centering
     \begin{subfigure}[h]{0.8\textwidth}
         \centering
         \includegraphics[width=\textwidth]{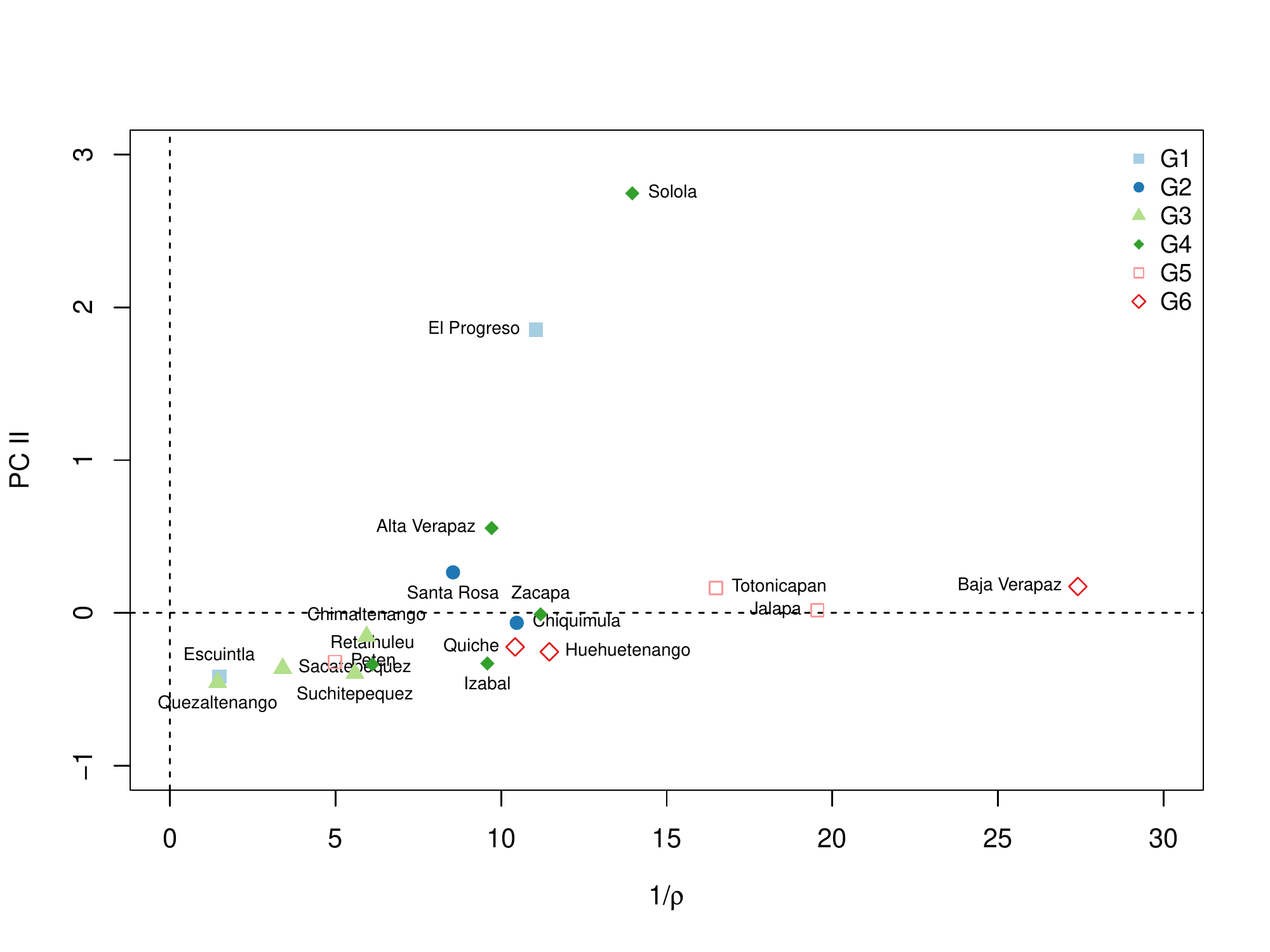}
         \subcaption{}
         \label{fig:5b}
     \end{subfigure}
     \caption{Relation between the second principal component $PCII$ and the importation risk (a), and the inverse virus importation risk (b).}\label{fig:5}
\end{figure}

Figures \ref{fig:2b} and \ref{fig:5} explore an hypothetical relationship between the principal component scores and the importation risk index $\rho$ defined above.  Figure \ref{fig:5a} shows a probable approximate relation between $PCII$ and the inverse of the importation risk index. In order to check this statement,  figure \ref{fig:5b} plots the second principal component $PCII$ as a function of $\rho^{-1}$. The figure suggests a positive correlation between $PCII$ and the inverse of the risk of importation index $\rho$. In the same figure, we show in different symbols and colors the groups $G1$ to $G6$ formed by the $PCI$ classification.

\begin{figure}
     \centering
         \includegraphics[width=0.8\textwidth]{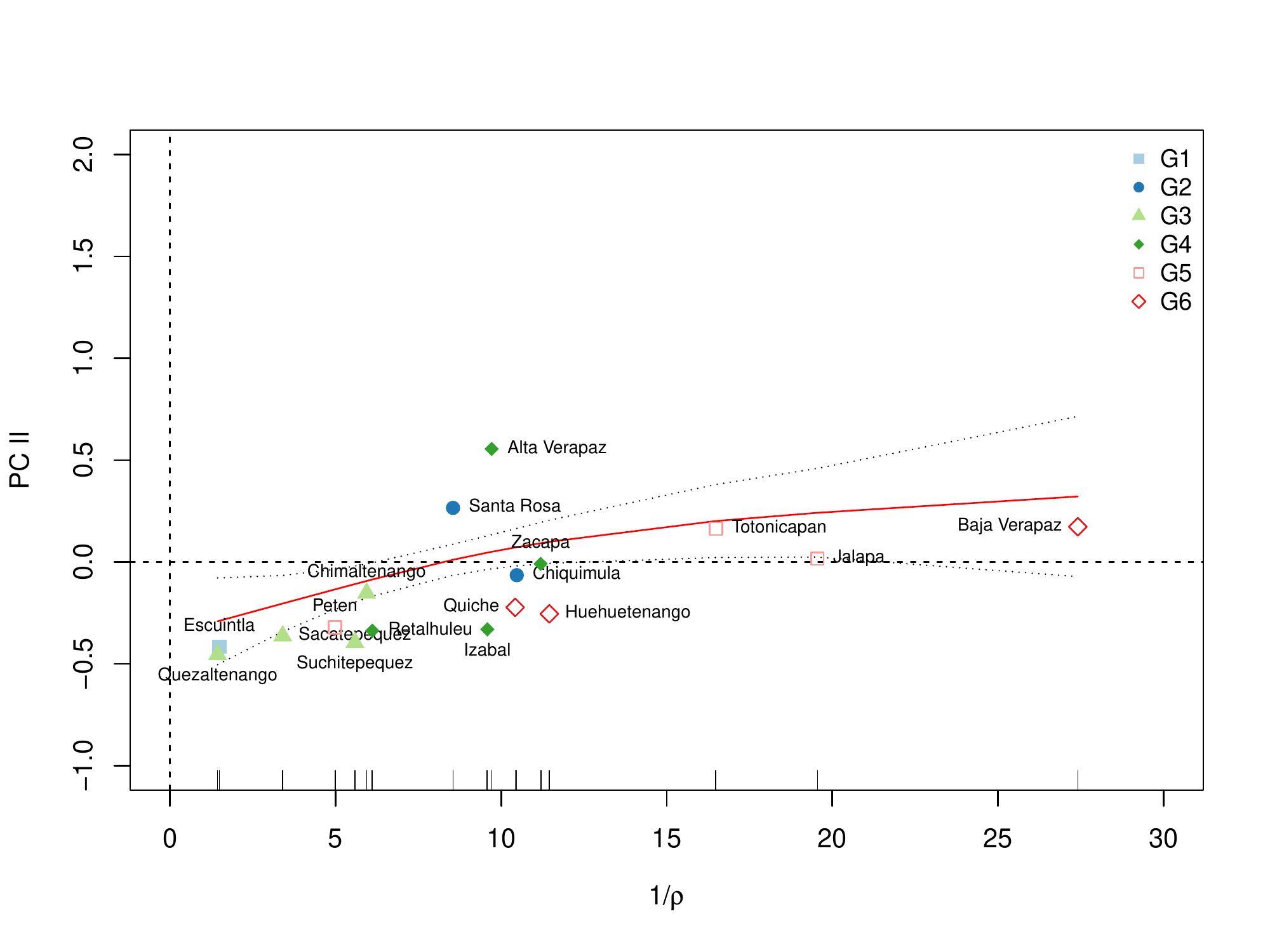}
\caption{ Generalized Additive Model fit }\label{fig:6}
\end{figure}

When we tested the correlation between $PCII$ scores and $\rho^{-1}$ using a linear model we found that the results are consistent with the hypothesis that the importation risk might be used as a covariable to explain the $PCII$ scores. Within the linear model, the estimated internal migratory flow data is correlated (Adjusted R-squared 0.3717, p-value 0.004676, non-parametric bootstrap \cite{Manly2018} 95\% confidence interval $(0.08,0.70)$), through the importation risk index, to the wave classification provided by the PCA technique.  Although the above results show a clear signal for a positive correlation, the relationship between $PCII$ and $\rho^{-1}$ is rather non-linear as observed in figures \ref{fig:5b} and \ref{fig:6}. 

One can clearly identify two departments, namely El Progreso ($PCII =1.85 $) and Solol\'a ($PCII =2.74$), which have exceptionally high $PCII$ scores compared to the rest of the data set.  Those departments have an importation risk  within the 1st quartile of the distribution of values.  El Progreso is also the less populated department in the country (176632 people) which can explain why its $PCII$ score, negatively correlated to the carrying capacity $K$ (see table \ref{table:2}), is high. On the other hand, the department of Solol\'a seems to be a special case. It has a population within the 2nd quartile (421583 people) and we cannot attempt such a straight explanation.

By relaxing the condition for a linear response and withdrawing the two above departments considered as outliers, we fit the resulting dataset to a Generalized Additive Model (GAM). The resulting value corresponding to deviance explained by the GAM, estimated against the null model is $47.3\%$. As a standard practice, this value is identified as a pseudo-$R^2$ for the model. Figure \ref{fig:6} shows the GAM fit to the data, together with the one-standard deviation region estimated by the model. 

Leaving aside the disease epicenters that we take as primary hot spots, Figure \ref{fig:5} demonstrates that the Escuintla department, another member of group $G1$, can be considered as a high-risk department for virus spread, given its $PCII$ score. Indeed, Table \ref{table:1} shows that the carrying capacity of Escuintla's first wave was relatively high, unlike departments such as El Progreso, which also belongs to $G1$ but has a lower carrying capacity (high $PCII$ score). We used this criterion to withdrawn El Progreso from the virus spread high-risk group. 
The Santa Rosa and Chiquimula departments, belonging to the $G2$ group, experienced an early peak within the initial COVID outbreak. However, they are characterized by intermediate $PCII$ and $\rho$ values, therefore they presumably do not contribute dominantly in shaping the spatial spread pattern. Therefore we'll consider them as belonging to a third category of risk.

Seemingly, a key role in epidemic spread is played by the $G3$ group, as its members have all similar wave patterns and relatively high importation risk index. Those localities have important urban centers which exhibited high community-level transmission during the pandemic and, additionally, present high connectivity with their nearby departments. According to the PCA scores, these departments can be considered as potential high-risk secondary cities in the epidemic outbreak pathway.

\begin{figure}
     \centering
         \includegraphics[width=\textwidth]{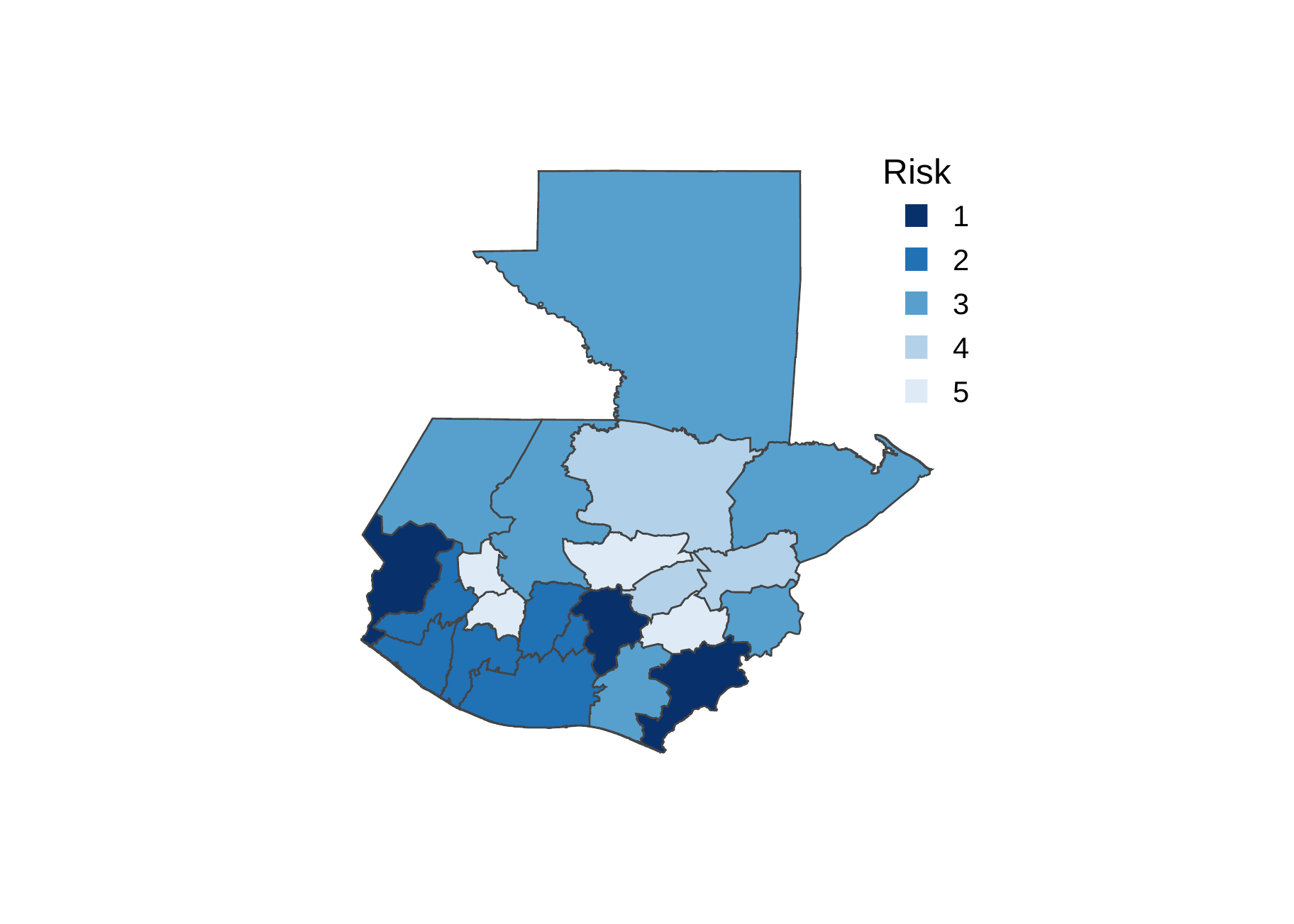}
\caption{Disease spread risk map of Guatemala based on COVID-19 initial wave outbreak}\label{fig:7}
     
\end{figure}

Finally, Figure \ref{fig:5b} strengthens the hypothesis that the departments from groups $G5$ and $G6$ have, on average, the lowest connectivity to the primary hot spots and therefore their epidemic wave peaks will appear delayed with respect to the first regional waves. We summarize the principal points of this discussion in the spread risk map of Figure \ref{fig:7}.

Heretofore, our findings strongly support employing the wave patterns landscape analysis as a guide to evaluate future epidemic spread risks in countries like Guatemala, which lack a strong epidemiological response data management infrastructure. 
 
\section{Conclusion}

\label{sec:Final}
In this work, we focused on the initial phase of the COVID-19 epidemic in Guatemala and used the regional growth curves characterization for spread risk assessment. Using the phenomenological Richards Model applied to the initial COVID-19 infection outbreak, we proposed an evidence-based risk map that highlights the regions which play a key role in the temporal progression  of the disease. The proposed risk map is grounded on a spatial association analysis, mainly provided by the classification of regional waves via the PCA technique and complemented by human mobility data.  The main contribution of the present work is the use of past epidemic growth profiles to reveal not only spatial similarities among different geographical regions but also previously unnoticed connections among these.  As a result, better predictions of an epidemiological process in spatial networks is obtained.

The utility of the proposed risk map goes beyond the development of the COVID-19 disease in Guatemala, and might be considered as a tool to guide intervention plans in future outbreaks of airborne communicable diseases with similar epidemiology. This is particularly relevant, considering that the early conditions of the COVID-19 epidemic progression experienced in Guatemala, where lax public health measures were adopted, are likely to be reproduced in new epidemic outbreaks. Further studies are required to better understand the coupled dynamics on networks. We believe that the elements presented here consist of an important insight and first step on top of which a thorough mechanistic modeling approach can be built.

\bmhead{Acknowledgments}

Funding for this study was provided by the International Development Research Centre (IDRC) and SICA-CSUCA for J.A.P., J.D.C., M.A and K.F.

\bibliography{covidGT-arXiv.bib}


\begin{thebibliography}{40}
\ifx \bisbn   \undefined \def \bisbn  #1{ISBN #1}\fi
\ifx \binits  \undefined \def \binits#1{#1}\fi
\ifx \bauthor  \undefined \def \bauthor#1{#1}\fi
\ifx \batitle  \undefined \def \batitle#1{#1}\fi
\ifx \bjtitle  \undefined \def \bjtitle#1{#1}\fi
\ifx \bvolume  \undefined \def \bvolume#1{\textbf{#1}}\fi
\ifx \byear  \undefined \def \byear#1{#1}\fi
\ifx \bissue  \undefined \def \bissue#1{#1}\fi
\ifx \bfpage  \undefined \def \bfpage#1{#1}\fi
\ifx \blpage  \undefined \def \blpage #1{#1}\fi
\ifx \burl  \undefined \def \burl#1{\textsf{#1}}\fi
\ifx \doiurl  \undefined \def \doiurl#1{\url{https://doi.org/#1}}\fi
\ifx \betal  \undefined \def \betal{\textit{et al.}}\fi
\ifx \binstitute  \undefined \def \binstitute#1{#1}\fi
\ifx \binstitutionaled  \undefined \def \binstitutionaled#1{#1}\fi
\ifx \bctitle  \undefined \def \bctitle#1{#1}\fi
\ifx \beditor  \undefined \def \beditor#1{#1}\fi
\ifx \bpublisher  \undefined \def \bpublisher#1{#1}\fi
\ifx \bbtitle  \undefined \def \bbtitle#1{#1}\fi
\ifx \bedition  \undefined \def \bedition#1{#1}\fi
\ifx \bseriesno  \undefined \def \bseriesno#1{#1}\fi
\ifx \blocation  \undefined \def \blocation#1{#1}\fi
\ifx \bsertitle  \undefined \def \bsertitle#1{#1}\fi
\ifx \bsnm \undefined \def \bsnm#1{#1}\fi
\ifx \bsuffix \undefined \def \bsuffix#1{#1}\fi
\ifx \bparticle \undefined \def \bparticle#1{#1}\fi
\ifx \barticle \undefined \def \barticle#1{#1}\fi
\bibcommenthead
\ifx \bconfdate \undefined \def \bconfdate #1{#1}\fi
\ifx \botherref \undefined \def \botherref #1{#1}\fi
\ifx \url \undefined \def \url#1{\textsf{#1}}\fi
\ifx \bchapter \undefined \def \bchapter#1{#1}\fi
\ifx \bbook \undefined \def \bbook#1{#1}\fi
\ifx \bcomment \undefined \def \bcomment#1{#1}\fi
\ifx \oauthor \undefined \def \oauthor#1{#1}\fi
\ifx \citeauthoryear \undefined \def \citeauthoryear#1{#1}\fi
\ifx \endbibitem  \undefined \def \endbibitem {}\fi
\ifx \bconflocation  \undefined \def \bconflocation#1{#1}\fi
\ifx \arxivurl  \undefined \def \arxivurl#1{\textsf{#1}}\fi
\csname PreBibitemsHook\endcsname

\bibitem{whoDashboard}
\begin{botherref}
\oauthor{\bsnm{{World Health Organization}}}:
{WHO Coronavirus (COVID-19) Dashboard}
(2022).
\url{https://covid19.who.int/}
\end{botherref}
\endbibitem

\bibitem{yang2020modified}
\begin{barticle}
\bauthor{\bsnm{Yang}, \binits{Z.}},
\bauthor{\bsnm{Zeng}, \binits{Z.}},
\bauthor{\bsnm{Wang}, \binits{K.}},
\bauthor{\bsnm{Wong}, \binits{S.S.}},
\bauthor{\bsnm{Liang}, \binits{W.}},
\bauthor{\bsnm{Zanin}, \binits{M.}},
\bauthor{\bsnm{Liu}, \binits{P.}},
\bauthor{\bsnm{Cao}, \binits{X.}},
\bauthor{\bsnm{Gao}, \binits{Z.}},
\bauthor{\bsnm{Mai}, \binits{Z.}},
\bauthor{\bsnm{Liang}, \binits{J.}},
\bauthor{\bsnm{Liu}, \binits{X.}},
\bauthor{\bsnm{Li}, \binits{S.}},
\bauthor{\bsnm{Li}, \binits{Y.}},
\bauthor{\bsnm{Ye}, \binits{F.}},
\bauthor{\bsnm{Guan}, \binits{W.}},
\bauthor{\bsnm{Yang}, \binits{Y.}},
\bauthor{\bsnm{Li}, \binits{F.}},
\bauthor{\bsnm{Luo}, \binits{S.}},
\bauthor{\bsnm{Xie}, \binits{Y.}},
\bauthor{\bsnm{Liu}, \binits{B.}},
\bauthor{\bsnm{Wang}, \binits{Z.}},
\bauthor{\bsnm{Zhang}, \binits{S.}},
\bauthor{\bsnm{Wang}, \binits{Y.}},
\bauthor{\bsnm{Zhong}, \binits{N.}},
\bauthor{\bsnm{He}, \binits{J.}}:
\batitle{{Modified SEIR and AI prediction of the epidemics trend of COVID-19 in
  China under public health interventions}}.
\bjtitle{Journal of Thoracic Disease}
\bvolume{12}(\bissue{3}),
\bfpage{165}--\blpage{174}
(\byear{2020}).
\doiurl{10.21037/jtd.2020.02.64}
\end{barticle}
\endbibitem

\bibitem{yan2020interpretable}
\begin{barticle}
\bauthor{\bsnm{Yan}, \binits{L.}},
\bauthor{\bsnm{Zhang}, \binits{H.-T.}},
\bauthor{\bsnm{Goncalves}, \binits{J.}},
\bauthor{\bsnm{Xiao}, \binits{Y.}},
\bauthor{\bsnm{Wang}, \binits{M.}},
\bauthor{\bsnm{Guo}, \binits{Y.}},
\bauthor{\bsnm{Sun}, \binits{C.}},
\bauthor{\bsnm{Tang}, \binits{X.}},
\bauthor{\bsnm{Jing}, \binits{L.}},
\bauthor{\bsnm{Zhang}, \binits{M.}},
\bauthor{\bsnm{Huang}, \binits{X.}},
\bauthor{\bsnm{Xiao}, \binits{Y.}},
\bauthor{\bsnm{Cao}, \binits{H.}},
\bauthor{\bsnm{Chen}, \binits{Y.}},
\bauthor{\bsnm{Ren}, \binits{T.}},
\bauthor{\bsnm{Wang}, \binits{F.}},
\bauthor{\bsnm{Xiao}, \binits{Y.}},
\bauthor{\bsnm{Huang}, \binits{S.}},
\bauthor{\bsnm{Tan}, \binits{X.}},
\bauthor{\bsnm{Huang}, \binits{N.}},
\bauthor{\bsnm{Jiao}, \binits{B.}},
\bauthor{\bsnm{Cheng}, \binits{C.}},
\bauthor{\bsnm{Zhang}, \binits{Y.}},
\bauthor{\bsnm{Luo}, \binits{A.}},
\bauthor{\bsnm{Mombaerts}, \binits{L.}},
\bauthor{\bsnm{Jin}, \binits{J.}},
\bauthor{\bsnm{Cao}, \binits{Z.}},
\bauthor{\bsnm{Li}, \binits{S.}},
\bauthor{\bsnm{Xu}, \binits{H.}},
\bauthor{\bsnm{Yuan}, \binits{Y.}}:
\batitle{{An interpretable mortality prediction model for COVID-19 patients}}.
\bjtitle{Nature Machine Intelligence}
\bvolume{2}(\bissue{5}),
\bfpage{283}--\blpage{288}
(\byear{2020}).
\doiurl{10.1038/s42256-020-0180-7}
\end{barticle}
\endbibitem

\bibitem{chatterjee2020147}
\begin{barticle}
\bauthor{\bsnm{Chatterjee}, \binits{K.}},
\bauthor{\bsnm{Chatterjee}, \binits{K.}},
\bauthor{\bsnm{Kumar}, \binits{A.}},
\bauthor{\bsnm{Shankar}, \binits{S.}}:
\batitle{{Healthcare impact of COVID-19 epidemic in India: A stochastic
  mathematical model}}.
\bjtitle{Medical Journal Armed Forces India}
\bvolume{76}(\bissue{2}),
\bfpage{147}--\blpage{155}
(\byear{2020}).
\doiurl{10.1016/j.mjafi.2020.03.022}
\end{barticle}
\endbibitem

\bibitem{Bertozzi16732}
\begin{barticle}
\bauthor{\bsnm{Bertozzi}, \binits{A.L.}},
\bauthor{\bsnm{Franco}, \binits{E.}},
\bauthor{\bsnm{Mohler}, \binits{G.}},
\bauthor{\bsnm{Short}, \binits{M.B.}},
\bauthor{\bsnm{Sledge}, \binits{D.}}:
\batitle{{The challenges of modeling and forecasting the spread of COVID-19}}.
\bjtitle{Proceedings of the National Academy of Sciences of the United States
  of America}
\bvolume{117}(\bissue{29}),
\bfpage{16732}--\blpage{16738}
(\byear{2020}).
\doiurl{10.1073/pnas.2006520117}
\end{barticle}
\endbibitem

\bibitem{shakeel2021covid}
\begin{barticle}
\bauthor{\bsnm{Shakeel}, \binits{S.M.}},
\bauthor{\bsnm{Kumar}, \binits{N.S.}},
\bauthor{\bsnm{Madalli}, \binits{P.P.}},
\bauthor{\bsnm{Srinivasaiah}, \binits{R.}},
\bauthor{\bsnm{Swamy}, \binits{D.R.}}:
\batitle{{Covid-19 prediction models: A systematic literature review}}.
\bjtitle{Osong Public Health and Research Perspectives}
\bvolume{12}(\bissue{4}),
\bfpage{215}--\blpage{229}
(\byear{2021}).
\doiurl{10.24171/J.PHRP.2021.0100}
\end{barticle}
\endbibitem

\bibitem{chang2021mobility}
\begin{barticle}
\bauthor{\bsnm{Chang}, \binits{S.}},
\bauthor{\bsnm{Pierson}, \binits{E.}},
\bauthor{\bsnm{Koh}, \binits{P.W.}},
\bauthor{\bsnm{Gerardin}, \binits{J.}},
\bauthor{\bsnm{Redbird}, \binits{B.}},
\bauthor{\bsnm{Grusky}, \binits{D.}},
\bauthor{\bsnm{Leskovec}, \binits{J.}}:
\batitle{{Mobility network models of COVID-19 explain inequities and inform
  reopening}}.
\bjtitle{Nature}
\bvolume{589}(\bissue{7840}),
\bfpage{82}--\blpage{87}
(\byear{2021}).
\doiurl{10.1038/s41586-020-2923-3}
\end{barticle}
\endbibitem

\bibitem{acuna2020108370}
\begin{barticle}
\bauthor{\bsnm{Acu{\~{n}}a-Zegarra}, \binits{M.A.}},
\bauthor{\bsnm{Santana-Cibrian}, \binits{M.}},
\bauthor{\bsnm{Velasco-Hernandez}, \binits{J.X.}}:
\batitle{{Modeling behavioral change and COVID-19 containment in Mexico: A
  trade-off between lockdown and compliance}}.
\bjtitle{Mathematical Biosciences}
\bvolume{325},
\bfpage{108370}
(\byear{2020}).
\doiurl{10.1016/j.mbs.2020.108370}
\end{barticle}
\endbibitem

\bibitem{lopez2020}
\begin{barticle}
\bauthor{\bsnm{L{\'{o}}pez}, \binits{L.}},
\bauthor{\bsnm{Rod{\'{o}}}, \binits{X.}}:
\batitle{{The end of social confinement and COVID-19 re-emergence risk}}.
\bjtitle{Nature Human Behaviour}
\bvolume{4}(\bissue{7}),
\bfpage{746}--\blpage{755}
(\byear{2020}).
\doiurl{10.1038/s41562-020-0908-8}
\end{barticle}
\endbibitem

\bibitem{jia2020population}
\begin{barticle}
\bauthor{\bsnm{Jia}, \binits{J.S.}},
\bauthor{\bsnm{Lu}, \binits{X.}},
\bauthor{\bsnm{Yuan}, \binits{Y.}},
\bauthor{\bsnm{Xu}, \binits{G.}},
\bauthor{\bsnm{Jia}, \binits{J.}},
\bauthor{\bsnm{Christakis}, \binits{N.A.}}:
\batitle{{Population flow drives spatio-temporal distribution of COVID-19 in
  China}}.
\bjtitle{Nature}
\bvolume{582}(\bissue{7812}),
\bfpage{389}--\blpage{394}
(\byear{2020}).
\doiurl{10.1038/s41586-020-2284-y}
\end{barticle}
\endbibitem

\bibitem{Li2021}
\begin{botherref}
\oauthor{\bsnm{Li}, \binits{Z.}},
\oauthor{\bsnm{Li}, \binits{H.}},
\oauthor{\bsnm{Zhang}, \binits{X.}},
\oauthor{\bsnm{Zhao}, \binits{C.}}:
{Estimation of human mobility patterns for forecasting the early spread of
  disease}.
Healthcare
\textbf{9}(9)
(2021).
\doiurl{10.3390/healthcare9091224}
\end{botherref}
\endbibitem

\bibitem{wu2020nowcasting}
\begin{barticle}
\bauthor{\bsnm{Wu}, \binits{J.T.}},
\bauthor{\bsnm{Leung}, \binits{K.}},
\bauthor{\bsnm{Leung}, \binits{G.M.}}:
\batitle{{Nowcasting and forecasting the potential domestic and international
  spread of the 2019-nCoV outbreak originating in Wuhan, China: a modelling
  study}}.
\bjtitle{The Lancet}
\bvolume{395}(\bissue{10225}),
\bfpage{689}--\blpage{697}
(\byear{2020}).
\doiurl{10.1016/S0140-6736(20)30260-9}
\end{barticle}
\endbibitem

\bibitem{richards1959flexible}
\begin{barticle}
\bauthor{\bsnm{Richards}, \binits{F.J.}}:
\batitle{{A Flexible Growth Function for Empirical Use}}.
\bjtitle{Journal of Experimental Botany}
\bvolume{10}(\bissue{29}),
\bfpage{290}--\blpage{300}
(\byear{1959})
\end{barticle}
\endbibitem

\bibitem{abdi2010principal}
\begin{barticle}
\bauthor{\bsnm{Abdi}, \binits{H.}},
\bauthor{\bsnm{Williams}, \binits{L.J.}}:
\batitle{{Principal component analysis}}.
\bjtitle{Wiley Interdisciplinary Reviews: Computational Statistics}
\bvolume{2}(\bissue{4}),
\bfpage{433}--\blpage{459}
(\byear{2010}).
\doiurl{10.1002/wics.101}
\end{barticle}
\endbibitem

\bibitem{ma2014estimating}
\begin{barticle}
\bauthor{\bsnm{Ma}, \binits{J.}},
\bauthor{\bsnm{Dushoff}, \binits{J.}},
\bauthor{\bsnm{Bolker}, \binits{B.M.}},
\bauthor{\bsnm{Earn}, \binits{D.J.D.}}:
\batitle{{Estimating Initial Epidemic Growth Rates}}.
\bjtitle{Bulletin of Mathematical Biology}
\bvolume{76}(\bissue{1}),
\bfpage{245}--\blpage{260}
(\byear{2014}).
\doiurl{10.1007/s11538-013-9918-2}
\end{barticle}
\endbibitem

\bibitem{chowell201666}
\begin{barticle}
\bauthor{\bsnm{Chowell}, \binits{G.}},
\bauthor{\bsnm{Sattenspiel}, \binits{L.}},
\bauthor{\bsnm{Bansal}, \binits{S.}},
\bauthor{\bsnm{Viboud}, \binits{C.}}:
\batitle{{Mathematical models to characterize early epidemic growth: A
  review}}.
\bjtitle{Physics of Life Reviews}
\bvolume{18},
\bfpage{66}--\blpage{97}
(\byear{2016}).
\doiurl{10.1016/j.plrev.2016.07.005}
\end{barticle}
\endbibitem

\bibitem{Pell2018}
\begin{barticle}
\bauthor{\bsnm{Pell}, \binits{B.}},
\bauthor{\bsnm{Kuang}, \binits{Y.}},
\bauthor{\bsnm{Viboud}, \binits{C.}},
\bauthor{\bsnm{Chowell}, \binits{G.}}:
\batitle{{Using phenomenological models for forecasting the 2015 Ebola
  challenge}}.
\bjtitle{Epidemics}
\bvolume{22},
\bfpage{62}--\blpage{70}
(\byear{2018}).
\doiurl{10.1016/j.epidem.2016.11.002}
\end{barticle}
\endbibitem

\bibitem{Brauer2019}
\begin{bbook}
\bauthor{\bsnm{Brauer}, \binits{F.}},
\bauthor{\bsnm{Castillo-Chavez}, \binits{C.}},
\bauthor{\bsnm{Feng}, \binits{Z.}}:
\bbtitle{Mathematical Models in Epidemiology},
pp. \bfpage{382}--\blpage{385}.
\bpublisher{Springer},
\blocation{New York}
(\byear{2019}).
\doiurl{10.1007/978-1-4939-9828-9\_17}
\end{bbook}
\endbibitem

\bibitem{viboud2016generalized}
\begin{barticle}
\bauthor{\bsnm{Viboud}, \binits{C.}},
\bauthor{\bsnm{Simonsen}, \binits{L.}},
\bauthor{\bsnm{Chowell}, \binits{G.}}:
\batitle{{A generalized-growth model to characterize the early ascending phase
  of infectious disease outbreaks}}.
\bjtitle{Epidemics}
\bvolume{15},
\bfpage{27}--\blpage{37}
(\byear{2016}).
\doiurl{10.1016/j.epidem.2016.01.002}
\end{barticle}
\endbibitem

\bibitem{heffernan2005}
\begin{barticle}
\bauthor{\bsnm{Heffernan}, \binits{J.M.}},
\bauthor{\bsnm{Smith}, \binits{R.J.}},
\bauthor{\bsnm{Wahl}, \binits{L.M.}}:
\batitle{Perspectives on the basic reproductive ratio}.
\bjtitle{Journal of the Royal Society Interface}
\bvolume{2}(\bissue{4}),
\bfpage{281}--\blpage{293}
(\byear{2005}).
\doiurl{10.1098/rsif.2005.0042}
\end{barticle}
\endbibitem

\bibitem{dennis1978analytical}
\begin{barticle}
\bauthor{\bsnm{Dennis}, \binits{B.}}:
\batitle{Analytical solution to an open-system model of population growth}.
\bjtitle{Mathematical Biosciences}
\bvolume{40}(\bissue{1-2}),
\bfpage{167}--\blpage{169}
(\byear{1978})
\end{barticle}
\endbibitem

\bibitem{wu2020generalized}
\begin{barticle}
\bauthor{\bsnm{Wu}, \binits{K.}},
\bauthor{\bsnm{Darcet}, \binits{D.}},
\bauthor{\bsnm{Wang}, \binits{Q.}},
\bauthor{\bsnm{Sornette}, \binits{D.}}:
\batitle{{Generalized logistic growth modeling of the COVID-19 outbreak:
  comparing the dynamics in the 29 provinces in China and in the rest of the
  world}}.
\bjtitle{Nonlinear Dynamics}
\bvolume{101}(\bissue{3}),
\bfpage{1561}--\blpage{1581}
(\byear{2020}).
\doiurl{10.1007/s11071-020-05862-6}
\end{barticle}
\endbibitem

\bibitem{sibly2005regulation}
\begin{barticle}
\bauthor{\bsnm{Sibly}, \binits{R.M.}},
\bauthor{\bsnm{Barker}, \binits{D.}},
\bauthor{\bsnm{Denham}, \binits{M.C.}},
\bauthor{\bsnm{Hone}, \binits{J.}},
\bauthor{\bsnm{Pagel}, \binits{M.}}:
\batitle{On the regulation of populations of mammals, birds, fish, and
  insects}.
\bjtitle{Science}
\bvolume{309}(\bissue{5734}),
\bfpage{607}--\blpage{610}
(\byear{2005})
\end{barticle}
\endbibitem

\bibitem{WANG201212}
\begin{barticle}
\bauthor{\bsnm{Wang}, \binits{X.S.}},
\bauthor{\bsnm{Wu}, \binits{J.}},
\bauthor{\bsnm{Yang}, \binits{Y.}}:
\batitle{{Richards model revisited: Validation by and application to infection
  dynamics}}.
\bjtitle{Journal of Theoretical Biology}
\bvolume{313},
\bfpage{12}--\blpage{19}
(\byear{2012}).
\doiurl{10.1016/j.jtbi.2012.07.024}
\end{barticle}
\endbibitem

\bibitem{Savitzky1964}
\begin{barticle}
\bauthor{\bsnm{Savitzky}, \binits{A.}},
\bauthor{\bsnm{Golay}, \binits{M.J.E.}}:
\batitle{{Smoothing and Differentiation of Data by Simplified Least Squares
  Procedure}}.
\bjtitle{Analytical Chemistry}
\bvolume{36}(\bissue{8}),
\bfpage{1627}--\blpage{1639}
(\byear{1964}).
\doiurl{10.1021/ac60214a047}
\end{barticle}
\endbibitem

\bibitem{johnson2014applied}
\begin{bbook}
\bauthor{\bsnm{Johnson}, \binits{R.A.}},
\bauthor{\bsnm{Wichern}, \binits{D.W.}}, \betal:
\bbtitle{Applied Multivariate Statistical Analysis}
vol. \bseriesno{6}.
\bpublisher{Pearson London, UK:}, \blocation{???}
(\byear{2014})
\end{bbook}
\endbibitem

\bibitem{INE2018}
\begin{botherref}
\oauthor{\bsnm{{National Institute of Statistics}}}:
{Cruce de variables}
(2019).
\url{http://redatam.censopoblacion.gt/bingtm/RpWebEngine.exe/Portal?BASE=CPVGT2018}
\end{botherref}
\endbibitem

\bibitem{Lai2020}
\begin{botherref}
\oauthor{\bsnm{Lai}, \binits{S.}},
\oauthor{\bsnm{Bogoch}, \binits{I.I.}},
\oauthor{\bsnm{Ruktanonchai}, \binits{N.W.}},
\oauthor{\bsnm{Watts}, \binits{A.}},
\oauthor{\bsnm{Lu}, \binits{X.}},
\oauthor{\bsnm{Yang}, \binits{W.}},
\oauthor{\bsnm{Yu}, \binits{H.}},
\oauthor{\bsnm{Khan}, \binits{K.}},
\oauthor{\bsnm{Tatem}, \binits{A.J.}}:
{Assessing spread risk of Wuhan novel coronavirus within and beyond China,
  January-April 2020: a travel network-based modelling study}
(2020).
\doiurl{10.1101/2020.02.04.20020479}
\end{botherref}
\endbibitem

\bibitem{Shanafelt2018}
\begin{barticle}
\bauthor{\bsnm{Shanafelt}, \binits{D.W.}},
\bauthor{\bsnm{Jones}, \binits{G.}},
\bauthor{\bsnm{Lima}, \binits{M.}},
\bauthor{\bsnm{Perrings}, \binits{C.}},
\bauthor{\bsnm{Chowell}, \binits{G.}}:
\batitle{{Forecasting the 2001 Foot-and-Mouth Disease Epidemic in the UK}}.
\bjtitle{EcoHealth}
\bvolume{15}(\bissue{2}),
\bfpage{338}--\blpage{347}
(\byear{2018}).
\doiurl{10.1007/s10393-017-1293-2}
\end{barticle}
\endbibitem

\bibitem{sanna2018spatial}
\begin{botherref}
\oauthor{\bsnm{Sanna}, \binits{M.}},
\oauthor{\bsnm{Wu}, \binits{J.}},
\oauthor{\bsnm{Zhu}, \binits{Y.}},
\oauthor{\bsnm{Yang}, \binits{Z.}},
\oauthor{\bsnm{Lu}, \binits{J.}},
\oauthor{\bsnm{Hsieh}, \binits{Y.H.}}:
{Spatial and Temporal Characteristics of 2014 Dengue Outbreak in Guangdong,
  China}.
Nature Scientific Reports
\textbf{8}
(2018).
\doiurl{10.1038/s41598-018-19168-6}
\end{botherref}
\endbibitem

\bibitem{Zuhairoh2020}
\begin{barticle}
\bauthor{\bsnm{Zuhairoh}, \binits{F.}},
\bauthor{\bsnm{Rosadi}, \binits{D.}}:
\batitle{{Real-time forecasting of the COVID-19 epidemic using the richards
  model in South Sulawesi, Indonesia}}.
\bjtitle{Indonesian Journal of Science and Technology}
\bvolume{5}(\bissue{3}),
\bfpage{456}--\blpage{462}
(\byear{2020}).
\doiurl{10.17509/ijost.v5i3.26139}
\end{barticle}
\endbibitem

\bibitem{Macedo2021}
\begin{barticle}
\bauthor{\bsnm{Mac{\^{e}}do}, \binits{A.M.S.}},
\bauthor{\bsnm{Brum}, \binits{A.A.}},
\bauthor{\bsnm{Duarte-Filho}, \binits{G.C.}},
\bauthor{\bsnm{Almeida}, \binits{F.A.G.}},
\bauthor{\bsnm{Ospina}, \binits{R.}},
\bauthor{\bsnm{Vasconcelos}, \binits{G.L.}}:
\batitle{{A Comparative Analysis between a SIRD Compartmental Model and the
  Richards Growth Model}}.
\bjtitle{Trends in Computational and Applied Mathematics}
\bvolume{22}(\bissue{4}),
\bfpage{545}--\blpage{557}
(\byear{2021}).
\doiurl{10.5540/tcam.2021.022.04.00545}
\end{barticle}
\endbibitem

\bibitem{hsieh2010epidemic}
\begin{barticle}
\bauthor{\bsnm{Hsieh}, \binits{Y.H.}},
\bauthor{\bsnm{Fisman}, \binits{D.N.}},
\bauthor{\bsnm{Wu}, \binits{J.}}:
\batitle{{On epidemic modeling in real time: An application to the 2009 Novel A
  (H1N1) influenza outbreak in Canada}}.
\bjtitle{BMC Research Notes}
\bvolume{3},
\bfpage{2}--\blpage{9}
(\byear{2010}).
\doiurl{10.1186/1756-0500-3-283}
\end{barticle}
\endbibitem

\bibitem{Markovic2021}
\begin{botherref}
\oauthor{\bsnm{Markovic}, \binits{S.}},
\oauthor{\bsnm{Rodic}, \binits{A.}},
\oauthor{\bsnm{Salom}, \binits{I.}},
\oauthor{\bsnm{Milicevic}, \binits{O.}},
\oauthor{\bsnm{Djordjevic}, \binits{M.}},
\oauthor{\bsnm{Djordjevic}, \binits{M.}}:
{COVID-19 severity determinants inferred through ecological and epidemiological
  modeling}.
One Health
\textbf{13}(August)
(2021).
\doiurl{10.1016/j.onehlt.2021.100355}
\end{botherref}
\endbibitem

\bibitem{Chen2021}
\begin{botherref}
\oauthor{\bsnm{Chen}, \binits{J.}},
\oauthor{\bsnm{Gong}, \binits{F.}},
\oauthor{\bsnm{Xiang}, \binits{S.}},
\oauthor{\bsnm{Yu}, \binits{T.}}:
{Application of principal component analysis in evaluation of epidemic
  situation policy implementation}.
Journal of Physics: Conference Series
\textbf{1903}
(2021).
\doiurl{10.1088/1742-6596/1903/1/012056}
\end{botherref}
\endbibitem

\bibitem{Agarwal2021}
\begin{barticle}
\bauthor{\bsnm{Agarwal}, \binits{D.P.}},
\bauthor{\bsnm{Ch}, \binits{M.A.}},
\bauthor{\bsnm{Kharate}, \binits{S.}},
\bauthor{\bsnm{Fantin}, \binits{E.}},
\bauthor{\bsnm{Raj}, \binits{I.}},
\bauthor{\bsnm{Balamuralitharan}, \binits{S.}}:
\batitle{Parameter estimation of covid-19 second wave b-h-r-p transmission
  model by using principle component analysis}.
\bjtitle{Annals of R.S.C.B.}
\bvolume{25}(\bissue{5}),
\bfpage{446}--\blpage{457}
(\byear{2021})
\end{barticle}
\endbibitem

\bibitem{Mahmoudi2021}
\begin{barticle}
\bauthor{\bsnm{Mahmoudi}, \binits{M.R.}},
\bauthor{\bsnm{Heydari}, \binits{M.H.}},
\bauthor{\bsnm{Qasem}, \binits{S.N.}},
\bauthor{\bsnm{Mosavi}, \binits{A.}},
\bauthor{\bsnm{Band}, \binits{S.S.}}:
\batitle{{Principal component analysis to study the relations between the
  spread rates of COVID-19 in high risks countries}}.
\bjtitle{Alexandria Engineering Journal}
\bvolume{60},
\bfpage{457}--\blpage{464}
(\byear{2021}).
\doiurl{10.1016/j.aej.2020.09.013}
\end{barticle}
\endbibitem

\bibitem{Gaudart2021}
\begin{barticle}
\bauthor{\bsnm{Gaudart}, \binits{J.}},
\bauthor{\bsnm{Landier}, \binits{J.}},
\bauthor{\bsnm{Huiart}, \binits{L.}},
\bauthor{\bsnm{Legendre}, \binits{E.}},
\bauthor{\bsnm{Lehot}, \binits{L.}},
\bauthor{\bsnm{Bendiane}, \binits{M.K.}},
\bauthor{\bsnm{Chiche}, \binits{L.}},
\bauthor{\bsnm{Petitjean}, \binits{A.}},
\bauthor{\bsnm{Mosnier}, \binits{E.}},
\bauthor{\bsnm{Kirakoya-Samadoulougou}, \binits{F.}},
\bauthor{\bsnm{Demongeot}, \binits{J.}},
\bauthor{\bsnm{Piarroux}, \binits{R.}},
\bauthor{\bsnm{Rebaudet}, \binits{S.}}:
\batitle{{Factors associated with the spatial heterogeneity of the first wave
  of COVID-19 in France: a nationwide geo-epidemiological study}}.
\bjtitle{The Lancet Public Health}
\bvolume{6}(\bissue{4}),
\bfpage{222}--\blpage{231}
(\byear{2021}).
\doiurl{10.1016/S2468-2667(21)00006-2}
\end{barticle}
\endbibitem

\bibitem{MSPAS}
\begin{botherref}
\oauthor{\bsnm{{Ministerio de Salud P{\'{u}}blica y Asistencia Social}}}:
{Situaci{\'{o}}n de COVID-19 en Guatemala}
(2022).
\url{https://tablerocovid.mspas.gob.gt/}
\end{botherref}
\endbibitem

\bibitem{Manly2018}
\begin{bbook}
\bauthor{\bsnm{Manly}, \binits{B.F.J.}}:
\bbtitle{Randomization, Bootstrap and Monte Carlo Methods in Biology}.
\bpublisher{Chapman and Hall/CRC}, \blocation{???}
(\byear{2018}).
\doiurl{10.1201/9781315273075}
\end{bbook}
\endbibitem

\end{thebibliography}

\end{document}